\documentclass[%
 aip,
 amsmath,amssymb,groupedaddress,
 reprint,%
]{revtex4-1}

\usepackage{graphicx}
\usepackage{dcolumn}
\usepackage{bm}

\usepackage[utf8]{inputenc}
\usepackage[T1]{fontenc}
\usepackage{mathptmx}
\usepackage{upgreek}
\usepackage{gensymb}
\usepackage{xcolor}
\usepackage{makecell}
\usepackage{nicefrac}

\usepackage{float}

\usepackage{titlesec}

\titleformat{\section}
{\normalfont\LARGE}{\thesection}{1em}{}

\titleformat{\subsection}
{\normalfont\Large}{\thesubsection.}{1em}{}

\newcommand{\beginsupplement}{%
	\setcounter{table}{0}
	\setcounter{figure}{0}
	\renewcommand{\thefigure}{S\arabic{figure}}%
}

\begin{document}

\preprint{AIP/123-QED}

\title{Free-standing silicon shadow masks for transmon qubit fabrication}

\author{I. Tsioutsios}
\email{ioannis.tsioutsios@yale.edu}
\affiliation{Department of Applied Physics, Yale University, New Haven, Connecticut 06520, USA}

\author{K. Serniak}%
\affiliation{Department of Applied Physics, Yale University, New Haven, Connecticut 06520, USA}

\author{S. Diamond}
\affiliation{Department of Applied Physics, Yale University, New Haven, Connecticut 06520, USA}

\author{V. V. Sivak}
\affiliation{Department of Applied Physics, Yale University, New Haven, Connecticut 06520, USA}

\author{Z. Wang}
\affiliation{Department of Applied Physics, Yale University, New Haven, Connecticut 06520, USA}

\author{S. Shankar}
\affiliation{Department of Applied Physics, Yale University, New Haven, Connecticut 06520, USA}

\author{L. Frunzio}
\affiliation{Department of Applied Physics, Yale University, New Haven, Connecticut 06520, USA}

\author{R. J. Schoelkopf}
\affiliation{Department of Applied Physics, Yale University, New Haven, Connecticut 06520, USA}

\author{M. H. Devoret}
\email{michel.devoret@yale.edu}
\affiliation{Department of Applied Physics, Yale University, New Haven, Connecticut 06520, USA}


\begin{abstract}
Nanofabrication techniques for superconducting qubits rely on resist-based masks patterned by electron-beam or optical lithography. We have developed an alternative nanofabrication technique based on free-standing silicon shadow masks fabricated from silicon-on-insulator wafers. These silicon shadow masks not only eliminate organic residues associated with resist-based lithography, but also provide a pathway to better understand and control surface-dielectric losses in superconducting qubits by decoupling mask fabrication from substrate preparation. We have successfully fabricated aluminum 3D transmon superconducting qubits with these shadow masks and found coherence quality factors comparable to those fabricated with standard techniques.
\end{abstract}

\maketitle

Progress in superconducting circuits for quantum information technologies relies on the improvement of superconducting qubit lifetimes\cite{Michel_Rob_science}. One of the main sources of energy loss in these devices comes from the dielectric surfaces surrounding the Josephson junctions and associated superconducting circuitry. In particular, a number of experimental results attribute the majority of dielectric loss to one or several of the device-substrate, substrate-air, and device-air interfaces, rather than the bulk dielectrics\cite{Geerlings2012,Oliver2013,Martinis2014,Quintana2014,Wang2015,Dial2016,Oliver2016,Gambetta2017,Calusine2018,Woods2019}.

State-of-the-art superconducting qubits are fabricated by patterning an organic resist with e-beam or optical lithography to create a liftoff mask, followed by shadow evaporation of the aluminum layer\cite{Dolan,Lecocq2011,Paik2011,Barends2013,Pop2014,Yan2015,Foroozani2019}. Inevitably, this approach introduces contamination to the various interfaces\cite{Quintana2014}. This includes organic residues from the resist, contamination from the solvents that are required for the resist development after e-beam exposure, and those required for the lift-off process after metal deposition. Furthermore, degassing of the organic mask during metal deposition can lead to additional contamination.

\begin{figure}
	\includegraphics{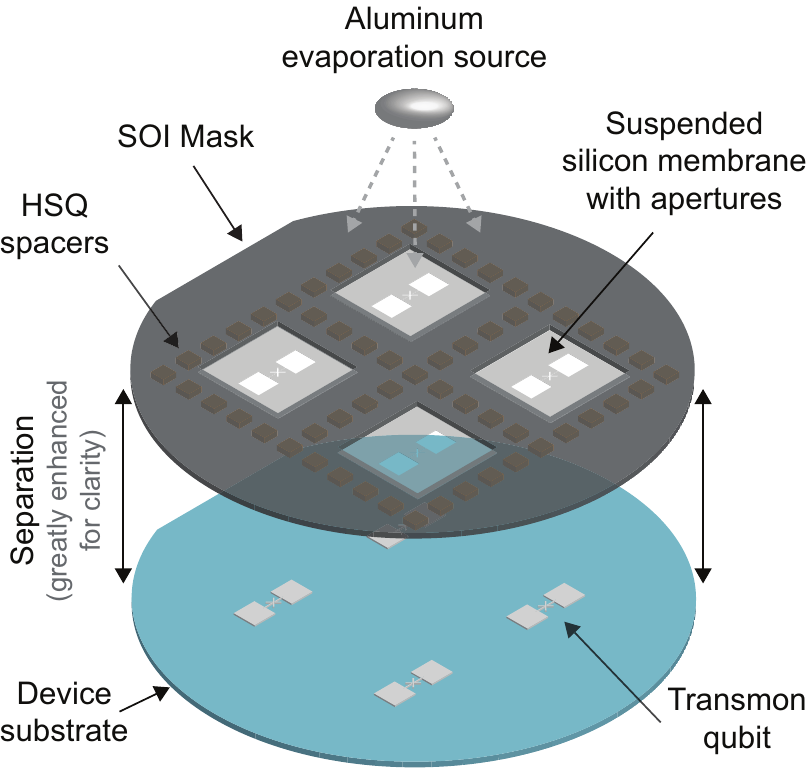}
	\caption{\label{figure_1} Concept for nanofabrication of superconducting transmon qubits using free-standing silicon shadow masks (not to scale illustration). A silicon-on-insulator (SOI) wafer incorporates micrometer-thick suspended silicon membranes, which contain apertures with submicron features. The stencil mask is placed on top of another wafer (device substrate). Aluminum is evaporated through it to create transmon structures on the device substrate. The micrometer-size cross-linked HSQ spacers control the distance between mask and device substrate. The mask is mechanically separated from the substrate at the end of aluminum deposition, leaving minimal nanofabrication-related residues. Here, the junction pattern has been caricatured for clarity.}
\end{figure}

In order to investigate the problems associated with residual contamination and eventually suppress it, we have developed a new nanofabrication technique for superconducting qubits (Fig.~\ref{figure_1}). Our technique replaces lift-off of an organic lithography layer with stencil lithography \cite{Vazquez2015} based on free-standing silicon shadow masks fabricated from silicon-on-insulator (SOI) wafers. Consequently, device substrate preparation becomes completely independent from the mask fabrication. As a result, the nanofabrication-related contamination is significantly reduced, and more importantly,  controlled studies of surface dielectric losses as a function of surface preparation are now possible. Moreover, the inorganic mask is compatible with high-temperature processes, such as deposition of refractory metals\cite{Place2020} and substrate annealing, which could be performed \textit{in situ}. The silicon mask is free-standing, and thus can be removed from the target substrate at the end of the process and reused for subsequent depositions. It is also tension-free and therefore has higher mechanical stability relative to other possible stencil methods.

\begin{figure}
	\includegraphics{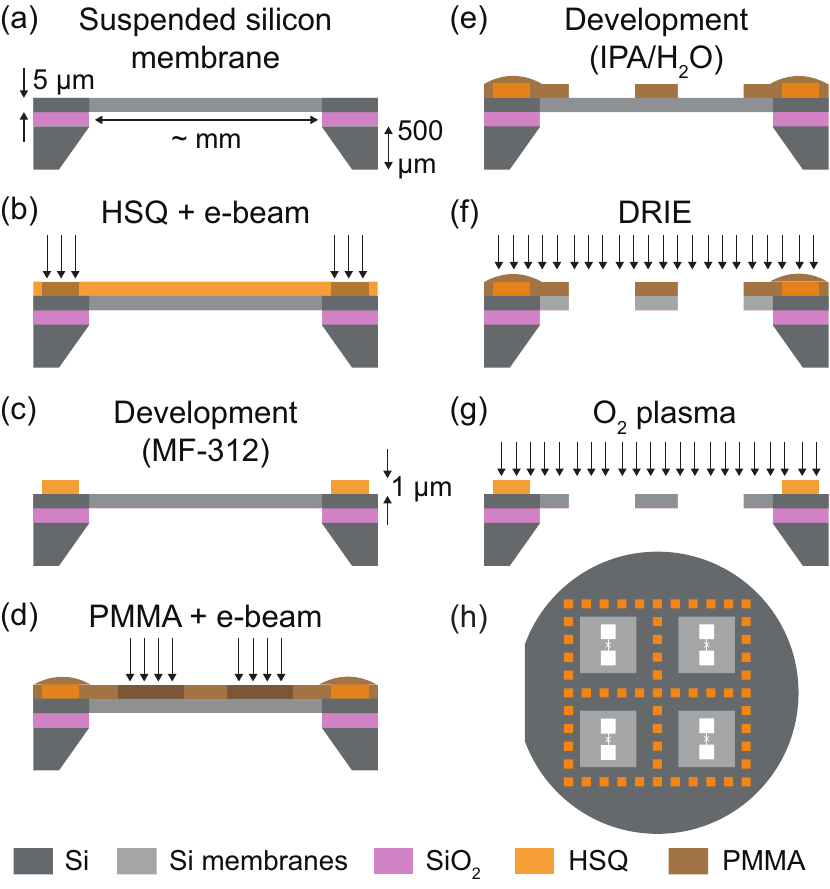}
	\caption{\label{figure_2} [(a)-(g)] Schematic cross section diagrams of the free-standing silicon shadow mask nanofabrication process (further described in the main text). (h) Top view schematic of the mask (not to scale).}
\end{figure}

The masks were fabricated from 100 mm SOI wafers, which consist of a $500 \ \mathrm{\upmu m}$-thick substrate, 200~nm-thick $\mathrm{SiO_{2}}$ layer and $5 \ \mathrm{\upmu m}$-thick silicon top layer. The wafers incorporate a prefabricated array of 60, $\mathrm{(2.7 \times 8.6) \ mm^{2}}$, $5 \ \mathrm{\upmu m}$-thick, suspended silicon membranes, where the silicon substrate and $\mathrm{SiO_{2}}$ layer were completely etched away\cite{Norcada}. A schematic cross section of a single suspended silicon membrane is illustrated in Fig.~\ref{figure_2}(a). The fabrication process starts by creating spacers to control the distance between mask and device substrate. The wafer was spin coated at 1000~rpm for 2~min with hydrogen silsesquioxane (HSQ - FOx16), which is a negative inorganic e-beam resist  [Fig.~\ref{figure_2}(b)]. It was then patterned in a Vistec electron-beam pattern generator (EBPG-5000+) with a 100 keV electron beam and developed in MF-312 for 5 minutes resulting in arrays of $(200 \times 200) \mathrm{\upmu m^{2}}$ and $1 \ \mathrm{\upmu m}$-thick cross-linked HSQ spacers [Figs.~\ref{figure_2}(c) and (h)]. Transmon patterns were defined by apertures in the silicon membranes, created with another step of e-beam lithography. The wafer was spin coated with PMMA 950 A7 resist at 1500~rpm, baked for 5~min at 200 $\mathrm{^{\circ}C}$, exposed with a 100 keV electron beam [Fig.~\ref{figure_2}(d)] and developed in $\mathrm{IPA/H_{2}O}$ (3:1) at 6 $\mathrm{^{\circ}C}$ for 2~min [Fig.~\ref{figure_2}(e)]. The apertures were created in the suspended silicon membranes by the highly anisotropic deep-reactive-ion-etching (DRIE) BOSCH process \cite{Bosch} [Fig.~\ref{figure_2}(f)]. As a last step, PMMA and other organic residues were removed from the mask with $\mathrm{O_2}$-plasma cleaning [Fig.~\ref{figure_2}(g)].

\begin{figure}
	\includegraphics{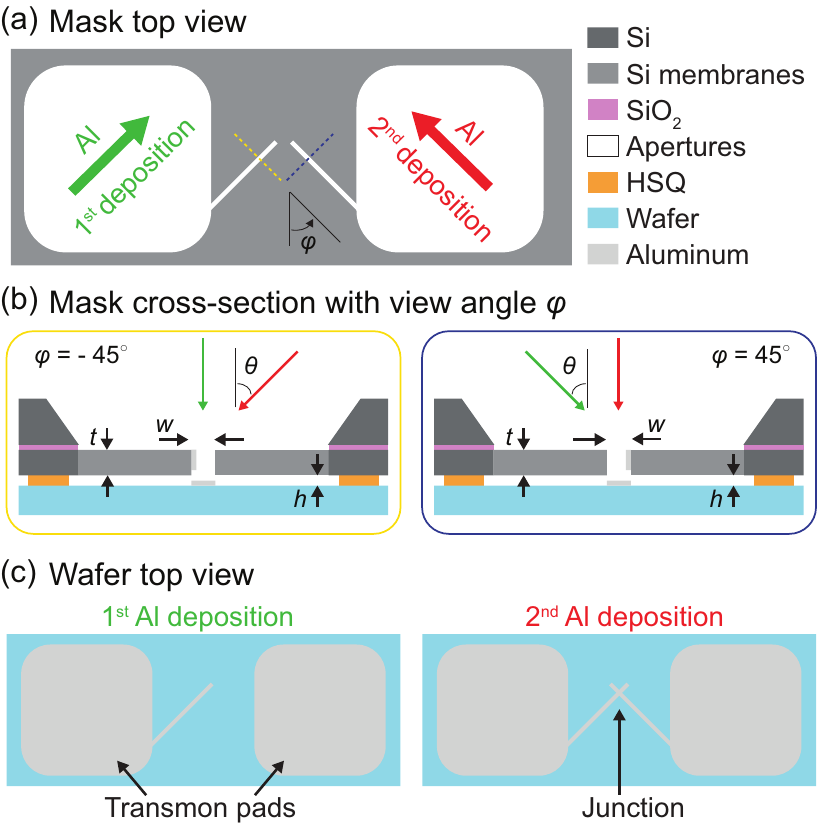}
	\caption{\label{figure_3} (a) Simplified mask design schematic. The large apertures correspond to the transmon qubit capacitor pads and the narrow slits to the leads of the Josephson tunnel junction. The green and red arrows indicate orientation of the Al deposition steps. (b) Schematic cross-sectional view (not to scale) of the mask at two distinct positions and angles indicated by the yellow and blue lines. Green and red arrows indicate the tilt angle $\theta$ of the first and second deposition steps, respectively. During the first deposition (green arrows), aluminum will only pass through the left narrow slit (yellow cross section), and it will land on the sidewalls of the other aperture (blue cross section). The reverse process occurs during the second deposition step (red arrows). (c) Top view schematic of the aluminum thin-film structure on the device wafer after each deposition step. The first Al deposition creates two capacitor pads and one thin lead, and the second Al deposition creates a second lead and contributes another layer to the capacitor pads. The Josephson junction is formed where these two leads cross.} 
\end{figure}

To demonstrate this new nanofabrication method, we focused on a mask design that is suitable for aluminum 3D transmon qubit\cite{Paik2011} fabrication. Figure~\ref{figure_3} is a simplified schematic describing the metal deposition method. The large rectangular apertures correspond to the capacitor pads and the narrow slits to the leads that will form the Josephson junction of the transmon. The deposition process requires the ability to tilt and rotate the mask-wafer stack  with respect to the evaporation source, similarly to that employed in the so-called ``Manhattan'' process\cite{Gladchenko2009}. The first deposition is performed with the stage rotated parallel to the left slit  $\mathrm{(\varphi=-45^\circ)}$ and tilted by angle $\mathrm{\theta}$, as shown in Figs.~\ref{figure_3}(a) and \ref{figure_3}(b) and determined by considerations below. By selecting the width of the junction slits to be much smaller than the thickness of the suspended silicon membranes, and selecting $\theta$ accordingly, aluminum is deposited through the left slit and lands on the sidewalls of the right slit [Fig.~\ref{figure_3}(b)]. To accomplish this, the minimum tilt angle should satisfy  $|\theta| > \arctan \left(\nicefrac{w}{t}\right)$, where $w$ is the width of the slit and $t$ the thickness of the silicon membrane. During the first deposition, the two capacitor pads and the first junction lead are formed, as shown in Fig.~\ref{figure_3}(c).  An \textit{in situ} oxidation step is then performed to create the tunnel barrier of the junction. A final (second) aluminum deposition with the stage rotated parallel to the right slit $\mathrm{(\varphi=45^\circ)}$ and tilted by $\mathrm{\theta}$ creates the second junction lead along with another aluminum layer on both capacitor pads [Fig.~\ref{figure_3}(c)].

\begin{figure}
	\includegraphics{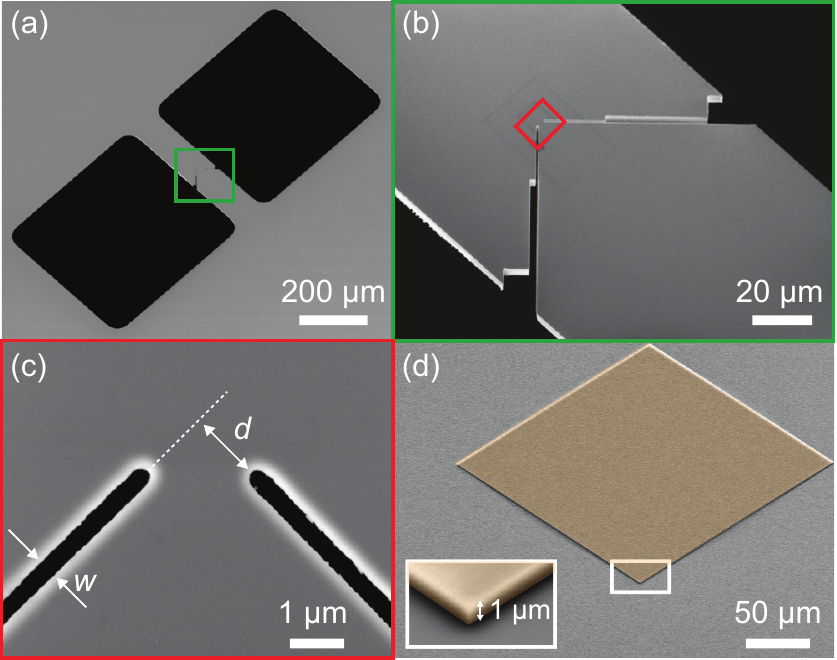}
	\caption{\label{figure_4} [(a)-(c)] Scanning electron microscopy (SEM) images of one of the silicon membranes of our free standing silicon shadow mask. Dark areas correspond to apertures and gray areas to suspended silicon. (d) False color SEM image of a $\mathrm{1 \ \upmu m}$ thick cross-linked HSQ spacer.}
\end{figure}

Each fabricated mask contains multiple suspended silicon membranes patterned in that way. In Figs.~\ref{figure_4}(a) - 4(c), scanning electron microscopy (SEM) images of a single silicon membrane of a mask are shown. In every membrane, the capacitor pad apertures have dimensions of $\mathrm{(530 \times 480) \ \upmu m^{2}}$. We designed the width of the junction lead slits such as it gradually reduces in order to minimize possible conductive losses from otherwise long and narrow aluminum leads [Fig.~\ref{figure_4}(b)]. We vary the minimum width of the junction lead slits $w$, from approximately $\mathrm{200}$ to $\mathrm{400~nm}$, in order to create transmons with different junction area from the same mask. Narrower slits would require further optimization of the DRIE process, as well as thinner silicon membranes\cite{Yeom2005}. In order to increase the mechanical stability of the suspended silicon structure after etching, we opted to end the lead slits well before their crossing point. This imposes an additional condition that the tilt angle satisfies $|\theta| > \arctan(\nicefrac{d}{h})$ for the two aluminum junction leads to overlap, where $h$ is the mask-substrate separation. The silicon membrane of $t=\mathrm{5 \ \upmu m}$ thickness provides the necessary bending rigidity which further increases the mechanical stability of the suspended structure. Much thinner silicon would require a modified mask design with in-plane bridges across the slits. In Fig.~\ref{figure_4}(d), the SEM image of a $\mathrm{(200 \times 200) \ \upmu m^{2}}$  and $\mathrm{1 \ \upmu m}$ thick cross-linked HSQ spacer is shown. Arrays of such spacers across the mask are meant to define $h$ and prevent possible adhesion of the mask on the device substrate due to van der Waals forces.

With the mask shown in Fig.~\ref{figure_4}, we fabricated arrays of 3D transmons\cite{Paik2011} on $200 \ \mathrm{\upmu m}$-thick, 100 mm diameter c-plane sapphire wafers. The sapphire substrates were cleaned in N-methyl-2-pyrrolidone (NMP) at $\mathrm{90 \ ^{\circ}C}$ for 10~min, sonicated consecutively in NMP, acetone, and isopropyl alcohol (IPA) for 3 min each, and then dried with nitrogen. All metal deposition and oxidation steps were performed in a Plassys UMS300UHV multichamber electron-beam evaporation system without breaking vacuum in-between steps. After reaching a base pressure less than $5\times10^{-9}$~Torr, we evaporated 21.3~nm aluminum at $\varphi=\mathrm{-45^{\circ}}$ and $\theta=\mathrm{20^{\circ}}$ at 1 nm/min rate. We then oxidized the aluminum \textit{in situ} with an $\mathrm{O_2/Ar}$ (3:17) mixture for 15 min at 100~Torr to create the tunnel barrier of the junction. A second evaporation of 31.9~nm aluminum was done at $\varphi=\mathrm{45^{\circ}}$ and $\theta=\mathrm{20^{\circ}}$. A final capping oxidation with an $\mathrm{O_2/Ar}$ (3:17) mixture for 5~min at 50~Torr was then performed. The same mask was employed multiple times on different sapphire wafers. The wafers were diced in $\mathrm{(8 \times 3) \ mm^{2}}$ chips, each containing a single transmon. To do so, we spin coated the wafers with a SC-1827 photoresist layer at 1500 rpm for 2~min and baked it at $90 \ ^{\circ}\mathrm{C}$ for 9 min.  This acts as protective layer against substrate debris damaging the devices during dicing.  The resist was stripped at the end of the dicing process using sequentially NMP, acetone, and IPA. Although the adoption of dicing resist is a convenient practice, it is in conflict here with one of the purposes of our proposed technique, which is to minimize fabrication residues, especially those coming from organic resist. However, the process of partitioning a wafer into smaller chips is independent of the fabrication of superconducting qubits at wafer-level, the main focus of our technique. The development of a reliable cleaving technique, which fundamentally does not require protective resist, would be essential for the full elimination of residues on the devices.  Nonetheless, acknowledging the above limitation, we tested these devices to determine whether our fabrication technique produces functional transmons.

\begin{figure}
	\includegraphics{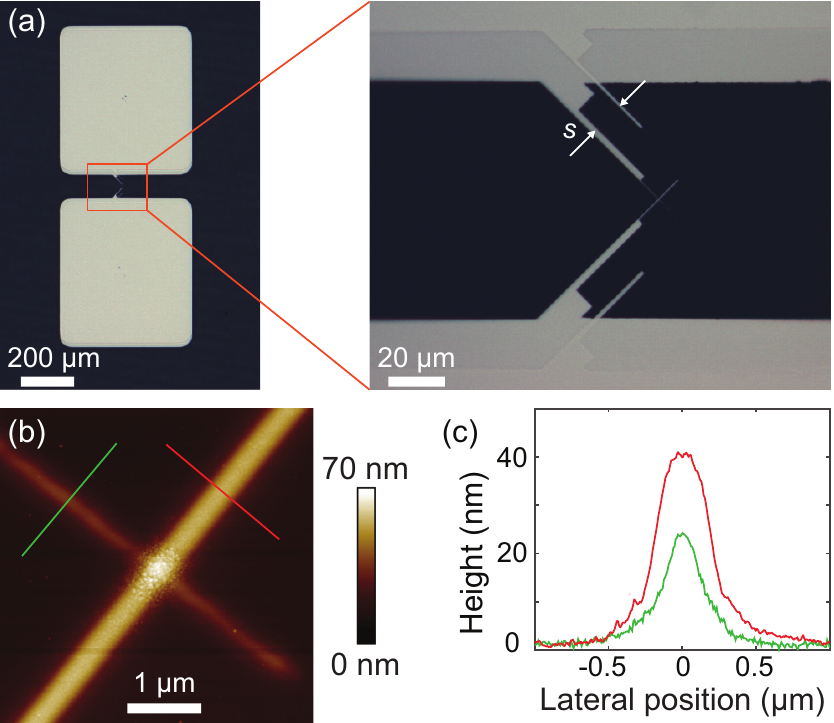}
	\caption{\label{figure_5} (a) Optical images of a transmon qubit device fabricated using a free-standing silicon shadow mask. The dark regions correspond to the sapphire substrate and the bright regions to deposited aluminum. (b) Atomic force microscopy (AFM) image of its Josephson junction formed at the crossing point of the two aluminum leads. (c) Height profile for each of the two aluminum leads at the cross sections indicated by green and red color lines in (b).}
\end{figure}

We characterized six of the fabricated aluminum transmon qubits, coming from two separate sapphire wafers, all corresponding to the mask shown in Fig.~\ref{figure_4}. In what follows, we present extensive results from one representative device, and partial results for the other five. In the optical images of Fig.~\ref{figure_5}(a), one can identify two aluminum layers that correspond to the two distinct evaporation steps. The double-strip pattern is expected for the double evaporation for a wide slit and does not affect the functionality of the devices. Nonetheless, the distance $s$ between the two strips provides an estimate for the effective mask-substrate separation of $h_\mathrm{eff}=s/\tan\theta = \mathrm{37 ~ \upmu m}$. This value is much larger than the thickness of the HSQ spacers ($\mathrm{1 \ \upmu m}$). We attribute this to built-in residual compressive strain in the silicon device layer of the SOI wafer \cite{Tsutomu2000}, which leads to buckling of the silicon membranes upon their release from the $\mathrm{Si/SiO_{2}}$ substrate. Nevertheless, a notable characteristic of our mask design is that the junction overlap area is approximately independent of the mask-substrate separation, as it is only defined by the width of the two slits. This contrasts with the results of the Dolan-bridge technique\cite{Dolan}, in which the junction area depends on both the mask substrate separation and the width of the slits. We further characterized the device by taking atomic force microscopy (AFM) images of its junction [Fig.~\ref{figure_5}(b)]. The asymmetry of the widths of the junction leads may be related to misalignment from the intended rotation angle $\varphi$, fabrication variances of the mask aperture widths, or more possibly due to different deposition thicknesses for each lead. A characteristic of this technique is that the deposited metallic thin films tend to have larger dimensions than the mask aperture sizes and softer edge profile [Fig.~\ref{figure_5}(c)], compared to traditional lift-off-based fabrication technologies. This blurring effect can have two distinct origins: the diffusion of the deposited material on the clean, resist-free surface of the substrate, and the geometry of the metal deposition process, which depends on the aluminum source size $D$, the source-mask seperation $H$, and the mask-substrate separation $h$\cite{Park2010,Vazquez2015}. The metal deposition setup that was used has a source size of $D = 5~\mathrm{mm}$ and a source-mask separation of $H = 0.6~\mathrm{m}$. For an effective mask-substrate separation of $h_\mathrm{eff} = \mathrm{37 ~ \upmu m}$, the enlargement of the feature sizes due to geometrical factors can be estimated to be $w_\mathrm{e}=Dh/H = 0.31~\mathrm{\upmu m}$. Knowing that the mask aperture widths that were used for this specific transmon were approximetely 200 nm, this second effect partly explains the aluminum leads profile of the device [Fig.~\ref{figure_5}(c)]. However, it cannot be the only feature enlargement effect, since the profile of the sidewalls is less steep than one would expect if the enlargement effect was only due to geometric factors. Therefore, we believe that the resulting soft and elongated cross-section profile of the aluminum junction leads [Fig.~\ref{figure_5}(c)] is a convolution of both diffusion and geometric effects. As blurring can be a limiting factor for smaller junction sizes, further investigation and modeling of the metal deposition dynamics is required. We further characterized the tunnel junction properties of the device, by measuring the normal-state resistance $R_\mathrm{n} = 6.9~\mathrm{K\Omega}$ of its tunnel junction, employing two-probe DC measurements, and estimating the critical current to be $I_\mathrm{c} = 41~\mathrm{nA}$ with the Ambegaokar–Baratoff formula $I_\mathrm{c}=(\pi \Delta)/(2eR_\mathrm{n})$, where $\Delta = 180 \mathrm{\ \upmu eV}$ is the aluminum superconducting gap. This value corresponds to a critical current density of $J_\mathrm{c} = 33~\mathrm{A/cm^2}$, assuming a junction area of $A_\mathrm{j} = 0.12~\upmu \mathrm{m^2}$, as extracted from the full-width-half-maximum of the junction lead profiles [Fig.~\ref{figure_5}(c)]. The average critical current densities that we measured for all the devices fabricated with the same mask on the same sapphire substrate is $\overline J_\mathrm{c}=42\pm 10~\mathrm{A/cm^2}$. This value is similar to those measured for usual aluminum tunnel junctions that we have fabricated with standard techniques and similar oxidation parameters.

\begin{figure}
	\includegraphics{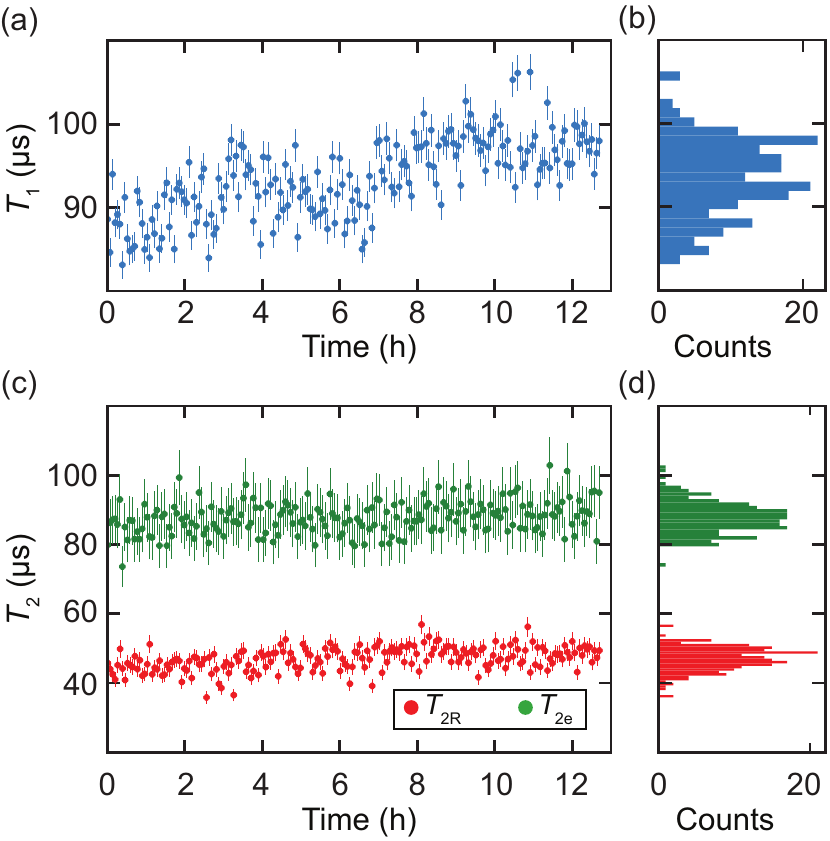}
	\caption{\label{figure_6} [(a) and (c)] Fluctuations of the coherence of the transmon qubit, which is shown in Fig.~\ref{figure_5}, as a function of time. Interleaved measurements of $T_\mathrm{1}$, $T_\mathrm{2R}$ Ramsey, and $T_\mathrm{2e}$ Hahn echo times were performed for approximately 13 h, with sampling times of 64 s, 101 s, and 65 s, respectively. The resulting mean measured values were $\overline T_1 \cong 95~\upmu \mathrm{s}$, $\overline T_\mathrm{2R} \cong 50~\upmu \mathrm{s}$, and $\overline T_\mathrm{2e} \cong 85~\upmu \mathrm{s}$. [(b) and (d)] Corresponding histograms for the measurements with standard deviations of $\sigma_{T_\mathrm{1}} \cong 5~\upmu \mathrm{s}$, $\sigma_{T_\mathrm{2R}} \cong 3~\upmu \mathrm{s}$, and $\sigma_{T_\mathrm{2e}} \cong 5~\upmu \mathrm{s}$.}
\end{figure}

The coherence properties of the transmon qubit, which is shown in Fig.~\ref{figure_5}, were measured in a dilution refrigerator with base temperature of approximately $\mathrm{20~mK}$, adopting a standard circuit quantum electrodynamics (cQED) architecture in the dispersive readout regime\cite{Blais2004}. The chip was mounted in an aluminum 3D rectangular-waveguide readout cavity\cite{Paik2011} with fundamental mode at frequency $\omega_\mathrm{r}/2\pi = \mathrm{9.1~GHz}$ (supplementary material). The measurements were performed in reflection and the input/output signals were coupled to the cavity though a single port with coupling rate set at $\kappa/2\pi=\mathrm{2.5~MHz}$. The reflected signal from the readout cavity was amplified by a near quantum limited Josephson array-mode parametric amplifier (JAMPA)\cite{Sivak2020}. The transmon had ground-to-first-excited-state transition frequency $\omega_\mathrm{ge}/2\pi=6.01~\mathrm{GHz}$, anharmonicity  $\alpha/2\pi=\omega_\mathrm{ge} - \omega_\mathrm{ef} = 0.23~\mathrm{GHz}$, and cross-Kerr to the readout cavity mode $\chi_\mathrm{qr} = 1.2~\mathrm{MHz}$. We characterized its coherence properties by performing interleaved repeated measurements of its $T_\mathrm{1}$ energy relaxation time, $T_\mathrm{2R}$ Ramsey and $T_\mathrm{2e}$ Hahn echo dephasing times for approximately 13 h [Fig.~\ref{figure_6}]. The length of each measurement was 64 s, 101 s, and 65 s, respectively. We found that the coherence values fluctuate in time with mean values of  $\overline T_1 \cong 95~\upmu \mathrm{s}$, $\overline T_\mathrm{2R}\cong50~\upmu \mathrm{s}$, $\overline T_\mathrm{2e}\cong85~\upmu \mathrm{s}$, and standard deviations of $\sigma_{T_\mathrm{1}} \cong 5~\upmu \mathrm{s}$, $\sigma_{T_\mathrm{2R}}\cong3~\upmu \mathrm{s}$, and $\sigma_{T_\mathrm{2e}} \cong 5~\upmu \mathrm{s}$. The measured $\overline T_\mathrm{1}$ corresponds to a quality factor of $\overline Q_\mathrm{ge} \cong 3.5\times10^6$, which is comparable to the state-of-the-art aluminum transmon qubits \cite{Kyle2019}. Nevertheless, further experimental studies are required to determine whether the energy relaxation properties of this device are limited by surface dielectric losses \cite{Wang2015}, nonequilibrium quasiparticle excitations\cite{Kyle2018,Kyle2019} or other loss mechanisms. The fluctuation of $T_\mathrm{1}$ and $T_\mathrm{2}$ as a function of time are similar to what we and other groups \cite{Paik2011,Klimov2018,Kyle2019} have observed with the same aluminum transmons but fabricated with standard techniques, and can be explained by uncontrolled sources of noise or fluctuating loss channels in the device. The low $T_\mathrm{{2R}}$ Ramsey and $T_\mathrm{{2e}}$ Hahn echo dephasing times, compared to the $T_\mathrm{2}=2T_\mathrm{1}$ limit, can be attributed to residual thermal photon population in the 3D aluminum readout cavity modes \cite{Zhixin2018}. The reproducibility of the fabrication technique was further assessed with coherence measurements of five more devices, fabricated with the same mask on two seperate sapphire wafers. All measurements yielded transmons with energy relaxation times of more than $16~\upmu \mathrm{s}$. A table with the full properties of all measured transmons that have been fabricated with the presented technique is given in the supplementary material.

Single tunnel junctions have been previously fabricated with free-standing shadow masks based on silicon nitride ($\mathrm{Si_3 N_4}$) membranes\cite{Ootuka1996,Savu2009}. However, in these efforts, the auxiliary probe-electrodes were fabricated in a separate step in advance. An advantage of free-standing membranes based on silicon, compared to $\mathrm{Si_3 Ni_4}$, is that they are nominally free from residual in-plane tensile stress. As a result, silicon masks are mechanically robust enough to implement complex asymmetric aperture designs, allowing for better control of the Josephson junction area independent of mask-substrate separation. Additionally, large and small features can coexist on the same membrane. This provided us the means to fabricate tunnel junctions and the necessary auxiliary circuitry of a superconducting qubit device, such as the large capacitor pads of a 3D transmon qubit, using a single free-standing mask, reducing fabrication residues on the entire qubit device. Furthermore, our technique eliminates the need to align the tunnel junction with respect to the auxiliary circuitry. Inorganic shadow mask based on Ge/Nb bilayer have also been used for the fabrication of aluminum tunnel junctions by Welander \textit{et al.}\cite{Welander2012}. In their work, Ge/Nb thin films are deposited and processed directly on the device substrate, which could potentially introduce additional contamination relative to free-standing inorganic masks.

In conclusion, we have developed a novel nanofabrication technique for superconducting qubits that is based on inorganic free-standing silicon shadow masks, fabricated from SOI wafers. We fabricated aluminum 3D transmon qubits with these masks and performed preliminary observations of their coherence properties. Our work addresses the residual contamination drawbacks inherent to e-beam and optical lithography techniques, providing a solid experimental platform to better understand, control, and potentially minimize surface-dielectric losses in planar superconducting circuits. This technique accomplishes full decoupling of the mask fabrication from device substrate preparation and thus minimizes cross-contamination between the mask and the device substrate. Systematic investigations of the effect of substrate treatment on surface dielectric losses without the restrictions imposed by organic resist processes are made possible. A key advantage of inorganic masks is their ability to sustain high metal deposition temperatures. To this end, free-standing silicon shadow masks hold promise as a suitable technique to fabricate high-quality superconducting qubits based on refractory materials with larger superconducting gap, such as niobium or tantalum\cite{Place2020}. In addition, high temperature substrate annealing\cite{Oliver2016} can now be achieved \textit{in situ}, under high vacuum, just before metal deposition, to further improve the surface properties of the device wafer. Finally, this technique is fully compatible with the fabrication of planar superconducting resonators, bringing to these necessary auxiliaries of tunnel junctions all of the aforementioned advantages.
\\


We acknowledge insightful discussions with Michael Rooks, Michael Power, Shantanu Mundhada, Chan U Lei and Andrew Saydjari. Facilities use was supported by YINQE and the Yale SEAS cleanroom. This research was supported by the Army Research Office (ARO) under Grant No. W911NF-18-1-0212, and Grant No. W911NF-18-1-0020, and by the Multidisciplinary University Research Initiatives-Office of Naval Research (MURI-ONR) under Grant No. N00014-16-1-2270.

\bibliography{FSSSMFTQF_v2.bib}

\begin{thebibliography}{35}%
\makeatletter
\providecommand \@ifxundefined [1]{%
 \@ifx{#1\undefined}
}%
\providecommand \@ifnum [1]{%
 \ifnum #1\expandafter \@firstoftwo
 \else \expandafter \@secondoftwo
 \fi
}%
\providecommand \@ifx [1]{%
 \ifx #1\expandafter \@firstoftwo
 \else \expandafter \@secondoftwo
 \fi
}%
\providecommand \natexlab [1]{#1}%
\providecommand \enquote  [1]{``#1''}%
\providecommand \bibnamefont  [1]{#1}%
\providecommand \bibfnamefont [1]{#1}%
\providecommand \citenamefont [1]{#1}%
\providecommand \href@noop [0]{\@secondoftwo}%
\providecommand \href [0]{\begingroup \@sanitize@url \@href}%
\providecommand \@href[1]{\@@startlink{#1}\@@href}%
\providecommand \@@href[1]{\endgroup#1\@@endlink}%
\providecommand \@sanitize@url [0]{\catcode `\\12\catcode `\$12\catcode
  `\&12\catcode `\#12\catcode `\^12\catcode `\_12\catcode `\%12\relax}%
\providecommand \@@startlink[1]{}%
\providecommand \@@endlink[0]{}%
\providecommand \url  [0]{\begingroup\@sanitize@url \@url }%
\providecommand \@url [1]{\endgroup\@href {#1}{\urlprefix }}%
\providecommand \urlprefix  [0]{URL }%
\providecommand \Eprint [0]{\href }%
\providecommand \doibase [0]{http://dx.doi.org/}%
\providecommand \selectlanguage [0]{\@gobble}%
\providecommand \bibinfo  [0]{\@secondoftwo}%
\providecommand \bibfield  [0]{\@secondoftwo}%
\providecommand \translation [1]{[#1]}%
\providecommand \BibitemOpen [0]{}%
\providecommand \bibitemStop [0]{}%
\providecommand \bibitemNoStop [0]{.\EOS\space}%
\providecommand \EOS [0]{\spacefactor3000\relax}%
\providecommand \BibitemShut  [1]{\csname bibitem#1\endcsname}%
\let\auto@bib@innerbib\@empty
\bibitem [{\citenamefont {Devoret}\ and\ \citenamefont
  {Schoelkopf}(2013)}]{Michel_Rob_science}%
  \BibitemOpen
  \bibfield  {author} {\bibinfo {author} {\bibfnamefont {M.~H.}\ \bibnamefont
  {Devoret}}\ and\ \bibinfo {author} {\bibfnamefont {R.~J.}\ \bibnamefont
  {Schoelkopf}},\ }\bibfield  {title} {\enquote {\bibinfo {title}
  {{Superconducting circuits for quantum information: an outlook}},}\ }\href
  {\doibase 10.1126/science.1231930} {\bibfield  {journal} {\bibinfo  {journal}
  {Science}\ }\textbf {\bibinfo {volume} {339}},\ \bibinfo {pages} {1169--1174}
  (\bibinfo {year} {2013})}\BibitemShut {NoStop}%
\bibitem [{\citenamefont {Geerlings}\ \emph {et~al.}(2012)\citenamefont
  {Geerlings}, \citenamefont {Shankar}, \citenamefont {Edwards}, \citenamefont
  {Frunzio}, \citenamefont {Schoelkopf},\ and\ \citenamefont
  {Devoret}}]{Geerlings2012}%
  \BibitemOpen
  \bibfield  {author} {\bibinfo {author} {\bibfnamefont {K.}~\bibnamefont
  {Geerlings}}, \bibinfo {author} {\bibfnamefont {S.}~\bibnamefont {Shankar}},
  \bibinfo {author} {\bibfnamefont {E.}~\bibnamefont {Edwards}}, \bibinfo
  {author} {\bibfnamefont {L.}~\bibnamefont {Frunzio}}, \bibinfo {author}
  {\bibfnamefont {R.~J.}\ \bibnamefont {Schoelkopf}}, \ and\ \bibinfo {author}
  {\bibfnamefont {M.~H.}\ \bibnamefont {Devoret}},\ }\bibfield  {title}
  {\enquote {\bibinfo {title} {{Improving the quality factor of microwave
  compact resonators by optimizing their geometrical parameters}},}\
  }\href@noop {} {\bibfield  {journal} {\bibinfo  {journal} {Applied Physics
  Letters}\ }\textbf {\bibinfo {volume} {100}},\ \bibinfo {pages} {192601}
  (\bibinfo {year} {2012})}\BibitemShut {NoStop}%
\bibitem [{\citenamefont {Oliver}\ and\ \citenamefont
  {Welander}(2013)}]{Oliver2013}%
  \BibitemOpen
  \bibfield  {author} {\bibinfo {author} {\bibfnamefont {W.~D.}\ \bibnamefont
  {Oliver}}\ and\ \bibinfo {author} {\bibfnamefont {P.~B.}\ \bibnamefont
  {Welander}},\ }\bibfield  {title} {\enquote {\bibinfo {title} {{Materials in
  superconducting quantum bits}},}\ }\href {\doibase 10.1557/mrs.2013.229}
  {\bibfield  {journal} {\bibinfo  {journal} {MRS Bull.}\ }\textbf {\bibinfo
  {volume} {38}},\ \bibinfo {pages} {816--825} (\bibinfo {year}
  {2013})}\BibitemShut {NoStop}%
\bibitem [{\citenamefont {Martinis}\ and\ \citenamefont
  {Megrant}(2014)}]{Martinis2014}%
  \BibitemOpen
  \bibfield  {author} {\bibinfo {author} {\bibfnamefont {J.~M.}\ \bibnamefont
  {Martinis}}\ and\ \bibinfo {author} {\bibfnamefont {A.}~\bibnamefont
  {Megrant}},\ }\bibfield  {title} {\enquote {\bibinfo {title} {{UCSB final
  report for the CSQ program: Review of decoherence and materials physics for
  superconducting qubits}},}\ }\href {http://arxiv.org/abs/1410.5793}
  {\bibfield  {journal} {\bibinfo  {journal} {arXiv}\ } (\bibinfo {year}
  {2014})},\ \Eprint {http://arxiv.org/abs/1410.5793v1} {1410.5793v1}
  \BibitemShut {NoStop}%
\bibitem [{\citenamefont {Quintana}\ \emph {et~al.}(2014)\citenamefont
  {Quintana}, \citenamefont {Megrant}, \citenamefont {Chen}, \citenamefont
  {Dunsworth}, \citenamefont {Chiaro}, \citenamefont {Barends}, \citenamefont
  {Campbell}, \citenamefont {Chen}, \citenamefont {Hoi}, \citenamefont
  {Jeffrey}, \citenamefont {Kelly}, \citenamefont {Mutus}, \citenamefont
  {O'Malley}, \citenamefont {Neill}, \citenamefont {Roushan}, \citenamefont
  {Sank}, \citenamefont {Vainsencher}, \citenamefont {Wenner}, \citenamefont
  {White}, \citenamefont {Cleland},\ and\ \citenamefont
  {Martinis}}]{Quintana2014}%
  \BibitemOpen
  \bibfield  {author} {\bibinfo {author} {\bibfnamefont {C.~M.}\ \bibnamefont
  {Quintana}}, \bibinfo {author} {\bibfnamefont {A.}~\bibnamefont {Megrant}},
  \bibinfo {author} {\bibfnamefont {Z.}~\bibnamefont {Chen}}, \bibinfo {author}
  {\bibfnamefont {A.}~\bibnamefont {Dunsworth}}, \bibinfo {author}
  {\bibfnamefont {B.}~\bibnamefont {Chiaro}}, \bibinfo {author} {\bibfnamefont
  {R.}~\bibnamefont {Barends}}, \bibinfo {author} {\bibfnamefont
  {B.}~\bibnamefont {Campbell}}, \bibinfo {author} {\bibfnamefont
  {Y.}~\bibnamefont {Chen}}, \bibinfo {author} {\bibfnamefont {I.~C.}\
  \bibnamefont {Hoi}}, \bibinfo {author} {\bibfnamefont {E.}~\bibnamefont
  {Jeffrey}}, \bibinfo {author} {\bibfnamefont {J.}~\bibnamefont {Kelly}},
  \bibinfo {author} {\bibfnamefont {J.~Y.}\ \bibnamefont {Mutus}}, \bibinfo
  {author} {\bibfnamefont {P.~J.~J.}\ \bibnamefont {O'Malley}}, \bibinfo
  {author} {\bibfnamefont {C.}~\bibnamefont {Neill}}, \bibinfo {author}
  {\bibfnamefont {P.}~\bibnamefont {Roushan}}, \bibinfo {author} {\bibfnamefont
  {D.}~\bibnamefont {Sank}}, \bibinfo {author} {\bibfnamefont {A.}~\bibnamefont
  {Vainsencher}}, \bibinfo {author} {\bibfnamefont {J.}~\bibnamefont {Wenner}},
  \bibinfo {author} {\bibfnamefont {T.~C.}\ \bibnamefont {White}}, \bibinfo
  {author} {\bibfnamefont {A.~N.}\ \bibnamefont {Cleland}}, \ and\ \bibinfo
  {author} {\bibfnamefont {J.~M.}\ \bibnamefont {Martinis}},\ }\bibfield
  {title} {\enquote {\bibinfo {title} {{Characterization and reduction of
  microfabrication-induced decoherence in superconducting quantum circuits}},}\
  }\href {\doibase 10.1063/1.4893297} {\bibfield  {journal} {\bibinfo
  {journal} {Applied Physics Letters}\ }\textbf {\bibinfo {volume} {105}},\
  \bibinfo {pages} {062601} (\bibinfo {year} {2014})}\BibitemShut {NoStop}%
\bibitem [{\citenamefont {Wang}\ \emph {et~al.}(2015)\citenamefont {Wang},
  \citenamefont {Axline}, \citenamefont {Gao}, \citenamefont {Brecht},
  \citenamefont {Chu}, \citenamefont {Frunzio}, \citenamefont {Devoret},\ and\
  \citenamefont {Schoelkopf}}]{Wang2015}%
  \BibitemOpen
  \bibfield  {author} {\bibinfo {author} {\bibfnamefont {C.}~\bibnamefont
  {Wang}}, \bibinfo {author} {\bibfnamefont {C.}~\bibnamefont {Axline}},
  \bibinfo {author} {\bibfnamefont {Y.~Y.}\ \bibnamefont {Gao}}, \bibinfo
  {author} {\bibfnamefont {T.}~\bibnamefont {Brecht}}, \bibinfo {author}
  {\bibfnamefont {Y.}~\bibnamefont {Chu}}, \bibinfo {author} {\bibfnamefont
  {L.}~\bibnamefont {Frunzio}}, \bibinfo {author} {\bibfnamefont {M.~H.}\
  \bibnamefont {Devoret}}, \ and\ \bibinfo {author} {\bibfnamefont {R.~J.}\
  \bibnamefont {Schoelkopf}},\ }\bibfield  {title} {\enquote {\bibinfo {title}
  {{Surface participation and dielectric loss in superconducting qubits}},}\
  }\href {\doibase 10.1063/1.4934486} {\bibfield  {journal} {\bibinfo
  {journal} {Applied Physics Letters}\ }\textbf {\bibinfo {volume} {107}},\
  \bibinfo {pages} {162601} (\bibinfo {year} {2015})}\BibitemShut {NoStop}%
\bibitem [{\citenamefont {Dial}\ \emph {et~al.}(2016)\citenamefont {Dial},
  \citenamefont {McClure}, \citenamefont {Poletto}, \citenamefont {Keefe},
  \citenamefont {Rothwell}, \citenamefont {Gambetta}, \citenamefont {Abraham},
  \citenamefont {Chow},\ and\ \citenamefont {Steffen}}]{Dial2016}%
  \BibitemOpen
  \bibfield  {author} {\bibinfo {author} {\bibfnamefont {O.}~\bibnamefont
  {Dial}}, \bibinfo {author} {\bibfnamefont {D.~T.}\ \bibnamefont {McClure}},
  \bibinfo {author} {\bibfnamefont {S.}~\bibnamefont {Poletto}}, \bibinfo
  {author} {\bibfnamefont {G.~A.}\ \bibnamefont {Keefe}}, \bibinfo {author}
  {\bibfnamefont {M.~B.}\ \bibnamefont {Rothwell}}, \bibinfo {author}
  {\bibfnamefont {J.~M.}\ \bibnamefont {Gambetta}}, \bibinfo {author}
  {\bibfnamefont {D.~W.}\ \bibnamefont {Abraham}}, \bibinfo {author}
  {\bibfnamefont {J.~M.}\ \bibnamefont {Chow}}, \ and\ \bibinfo {author}
  {\bibfnamefont {M.}~\bibnamefont {Steffen}},\ }\bibfield  {title} {\enquote
  {\bibinfo {title} {{Bulk and surface loss in superconducting transmon
  qubits}},}\ }\href {\doibase 10.1088/0953-2048/29/4/044001} {\bibfield
  {journal} {\bibinfo  {journal} {Superconductor Science and Technology}\
  }\textbf {\bibinfo {volume} {29}},\ \bibinfo {pages} {044001} (\bibinfo
  {year} {2016})}\BibitemShut {NoStop}%
\bibitem [{\citenamefont {Kamal}\ \emph {et~al.}(2016)\citenamefont {Kamal},
  \citenamefont {Yoder}, \citenamefont {Yan}, \citenamefont {Gudmundsen},
  \citenamefont {Hover}, \citenamefont {Sears}, \citenamefont {Welander},
  \citenamefont {Orlando}, \citenamefont {Gustavsson},\ and\ \citenamefont
  {Oliver}}]{Oliver2016}%
  \BibitemOpen
  \bibfield  {author} {\bibinfo {author} {\bibfnamefont {A.}~\bibnamefont
  {Kamal}}, \bibinfo {author} {\bibfnamefont {J.~L.}\ \bibnamefont {Yoder}},
  \bibinfo {author} {\bibfnamefont {F.}~\bibnamefont {Yan}}, \bibinfo {author}
  {\bibfnamefont {T.~J.}\ \bibnamefont {Gudmundsen}}, \bibinfo {author}
  {\bibfnamefont {D.}~\bibnamefont {Hover}}, \bibinfo {author} {\bibfnamefont
  {A.~P.}\ \bibnamefont {Sears}}, \bibinfo {author} {\bibfnamefont
  {P.}~\bibnamefont {Welander}}, \bibinfo {author} {\bibfnamefont {T.~P.}\
  \bibnamefont {Orlando}}, \bibinfo {author} {\bibfnamefont {S.}~\bibnamefont
  {Gustavsson}}, \ and\ \bibinfo {author} {\bibfnamefont {W.~D.}\ \bibnamefont
  {Oliver}},\ }\bibfield  {title} {\enquote {\bibinfo {title} {{Improved
  superconducting qubit coherence with high-temperature substrate
  annealing}},}\ }\href {http://arxiv.org/abs/1606.09262} {\  (\bibinfo {year}
  {2016})},\ \Eprint {http://arxiv.org/abs/1606.09262v1} {arXiv:1606.09262v1}
  \BibitemShut {NoStop}%
\bibitem [{\citenamefont {Gambetta}\ \emph {et~al.}(2017)\citenamefont
  {Gambetta}, \citenamefont {Murray}, \citenamefont {Fung}, \citenamefont
  {Mcclure}, \citenamefont {Dial}, \citenamefont {Shanks}, \citenamefont
  {Sleight},\ and\ \citenamefont {Steffen}}]{Gambetta2017}%
  \BibitemOpen
  \bibfield  {author} {\bibinfo {author} {\bibfnamefont {J.~M.}\ \bibnamefont
  {Gambetta}}, \bibinfo {author} {\bibfnamefont {C.~E.}\ \bibnamefont
  {Murray}}, \bibinfo {author} {\bibfnamefont {Y.~K.}\ \bibnamefont {Fung}},
  \bibinfo {author} {\bibfnamefont {D.~T.}\ \bibnamefont {Mcclure}}, \bibinfo
  {author} {\bibfnamefont {O.}~\bibnamefont {Dial}}, \bibinfo {author}
  {\bibfnamefont {W.}~\bibnamefont {Shanks}}, \bibinfo {author} {\bibfnamefont
  {J.~W.}\ \bibnamefont {Sleight}}, \ and\ \bibinfo {author} {\bibfnamefont
  {M.}~\bibnamefont {Steffen}},\ }\bibfield  {title} {\enquote {\bibinfo
  {title} {{Investigating surface loss effects in superconducting transmon
  qubits}},}\ }\href {\doibase 10.1109/TASC.2016.2629670} {\bibfield  {journal}
  {\bibinfo  {journal} {IEEE Transactions on Applied Superconductivity}\
  }\textbf {\bibinfo {volume} {27}},\ \bibinfo {pages} {1--5} (\bibinfo {year}
  {2017})}\BibitemShut {NoStop}%
\bibitem [{\citenamefont {Calusine}\ \emph {et~al.}(2018)\citenamefont
  {Calusine}, \citenamefont {Melville}, \citenamefont {Woods}, \citenamefont
  {Das}, \citenamefont {Stull}, \citenamefont {Bolkhovsky}, \citenamefont
  {Braje}, \citenamefont {Hover}, \citenamefont {Kim}, \citenamefont {Miloshi},
  \citenamefont {Rosenberg}, \citenamefont {Sevi}, \citenamefont {Yoder},
  \citenamefont {Dauler},\ and\ \citenamefont {Oliver}}]{Calusine2018}%
  \BibitemOpen
  \bibfield  {author} {\bibinfo {author} {\bibfnamefont {G.}~\bibnamefont
  {Calusine}}, \bibinfo {author} {\bibfnamefont {A.}~\bibnamefont {Melville}},
  \bibinfo {author} {\bibfnamefont {W.}~\bibnamefont {Woods}}, \bibinfo
  {author} {\bibfnamefont {R.}~\bibnamefont {Das}}, \bibinfo {author}
  {\bibfnamefont {C.}~\bibnamefont {Stull}}, \bibinfo {author} {\bibfnamefont
  {V.}~\bibnamefont {Bolkhovsky}}, \bibinfo {author} {\bibfnamefont
  {D.}~\bibnamefont {Braje}}, \bibinfo {author} {\bibfnamefont
  {D.}~\bibnamefont {Hover}}, \bibinfo {author} {\bibfnamefont {D.~K.}\
  \bibnamefont {Kim}}, \bibinfo {author} {\bibfnamefont {X.}~\bibnamefont
  {Miloshi}}, \bibinfo {author} {\bibfnamefont {D.}~\bibnamefont {Rosenberg}},
  \bibinfo {author} {\bibfnamefont {A.}~\bibnamefont {Sevi}}, \bibinfo {author}
  {\bibfnamefont {J.~L.}\ \bibnamefont {Yoder}}, \bibinfo {author}
  {\bibfnamefont {E.}~\bibnamefont {Dauler}}, \ and\ \bibinfo {author}
  {\bibfnamefont {W.~D.}\ \bibnamefont {Oliver}},\ }\bibfield  {title}
  {\enquote {\bibinfo {title} {{Analysis and mitigation of interface losses in
  trenched superconducting coplanar waveguide resonators}},}\ }\href {\doibase
  10.1063/1.5006888} {\bibfield  {journal} {\bibinfo  {journal} {Applied
  Physics Letters}\ }\textbf {\bibinfo {volume} {112}},\ \bibinfo {pages}
  {062601} (\bibinfo {year} {2018})}\BibitemShut {NoStop}%
\bibitem [{\citenamefont {Woods}\ \emph {et~al.}(2019)\citenamefont {Woods},
  \citenamefont {Calusine}, \citenamefont {Melville}, \citenamefont {Sevi},
  \citenamefont {Golden}, \citenamefont {Kim}, \citenamefont {Rosenberg},
  \citenamefont {Yoder},\ and\ \citenamefont {Oliver}}]{Woods2019}%
  \BibitemOpen
  \bibfield  {author} {\bibinfo {author} {\bibfnamefont {W.}~\bibnamefont
  {Woods}}, \bibinfo {author} {\bibfnamefont {G.}~\bibnamefont {Calusine}},
  \bibinfo {author} {\bibfnamefont {A.}~\bibnamefont {Melville}}, \bibinfo
  {author} {\bibfnamefont {A.}~\bibnamefont {Sevi}}, \bibinfo {author}
  {\bibfnamefont {E.}~\bibnamefont {Golden}}, \bibinfo {author} {\bibfnamefont
  {D.}~\bibnamefont {Kim}}, \bibinfo {author} {\bibfnamefont {D.}~\bibnamefont
  {Rosenberg}}, \bibinfo {author} {\bibfnamefont {J.}~\bibnamefont {Yoder}}, \
  and\ \bibinfo {author} {\bibfnamefont {W.}~\bibnamefont {Oliver}},\
  }\bibfield  {title} {\enquote {\bibinfo {title} {{Determining Interface
  Dielectric Losses in Superconducting Coplanar-Waveguide Resonators}},}\
  }\href {\doibase 10.1103/PhysRevApplied.12.014012} {\bibfield  {journal}
  {\bibinfo  {journal} {Physical Review Applied}\ }\textbf {\bibinfo {volume}
  {12}},\ \bibinfo {pages} {014012} (\bibinfo {year} {2019})}\BibitemShut
  {NoStop}%
\bibitem [{\citenamefont {Dolan}(1977)}]{Dolan}%
  \BibitemOpen
  \bibfield  {author} {\bibinfo {author} {\bibfnamefont {G.~J.}\ \bibnamefont
  {Dolan}},\ }\bibfield  {title} {\enquote {\bibinfo {title} {{Offset masks for
  lift-off photoprocessing}},}\ }\href {\doibase 10.1063/1.89690} {\bibfield
  {journal} {\bibinfo  {journal} {Applied Physics Letters}\ }\textbf {\bibinfo
  {volume} {31}},\ \bibinfo {pages} {337--339} (\bibinfo {year}
  {1977})}\BibitemShut {NoStop}%
\bibitem [{\citenamefont {Lecocq}\ \emph {et~al.}(2011)\citenamefont {Lecocq},
  \citenamefont {Pop}, \citenamefont {Peng}, \citenamefont {Matei},
  \citenamefont {Crozes}, \citenamefont {Fournier}, \citenamefont {Naud},
  \citenamefont {Guichard},\ and\ \citenamefont {Buisson}}]{Lecocq2011}%
  \BibitemOpen
  \bibfield  {author} {\bibinfo {author} {\bibfnamefont {F.}~\bibnamefont
  {Lecocq}}, \bibinfo {author} {\bibfnamefont {I.~M.}\ \bibnamefont {Pop}},
  \bibinfo {author} {\bibfnamefont {Z.}~\bibnamefont {Peng}}, \bibinfo {author}
  {\bibfnamefont {I.}~\bibnamefont {Matei}}, \bibinfo {author} {\bibfnamefont
  {T.}~\bibnamefont {Crozes}}, \bibinfo {author} {\bibfnamefont
  {T.}~\bibnamefont {Fournier}}, \bibinfo {author} {\bibfnamefont
  {C.}~\bibnamefont {Naud}}, \bibinfo {author} {\bibfnamefont {W.}~\bibnamefont
  {Guichard}}, \ and\ \bibinfo {author} {\bibfnamefont {O.}~\bibnamefont
  {Buisson}},\ }\bibfield  {title} {\enquote {\bibinfo {title} {{Junction
  fabrication by shadow evaporation without a suspended bridge}},}\ }\href@noop
  {} {\bibfield  {journal} {\bibinfo  {journal} {Nanotechnology}\ }\textbf
  {\bibinfo {volume} {22}},\ \bibinfo {pages} {315302} (\bibinfo {year}
  {2011})}\BibitemShut {NoStop}%
\bibitem [{\citenamefont {Paik}\ \emph {et~al.}(2011)\citenamefont {Paik},
  \citenamefont {Schuster}, \citenamefont {Bishop}, \citenamefont {Kirchmair},
  \citenamefont {Catelani}, \citenamefont {Sears}, \citenamefont {Johnson},
  \citenamefont {Reagor}, \citenamefont {Frunzio}, \citenamefont {Glazman},
  \citenamefont {Girvin}, \citenamefont {Devoret},\ and\ \citenamefont
  {Schoelkopf}}]{Paik2011}%
  \BibitemOpen
  \bibfield  {author} {\bibinfo {author} {\bibfnamefont {H.}~\bibnamefont
  {Paik}}, \bibinfo {author} {\bibfnamefont {D.~I.}\ \bibnamefont {Schuster}},
  \bibinfo {author} {\bibfnamefont {L.~S.}\ \bibnamefont {Bishop}}, \bibinfo
  {author} {\bibfnamefont {G.}~\bibnamefont {Kirchmair}}, \bibinfo {author}
  {\bibfnamefont {G.}~\bibnamefont {Catelani}}, \bibinfo {author}
  {\bibfnamefont {A.~P.}\ \bibnamefont {Sears}}, \bibinfo {author}
  {\bibfnamefont {B.~R.}\ \bibnamefont {Johnson}}, \bibinfo {author}
  {\bibfnamefont {M.~J.}\ \bibnamefont {Reagor}}, \bibinfo {author}
  {\bibfnamefont {L.}~\bibnamefont {Frunzio}}, \bibinfo {author} {\bibfnamefont
  {L.~I.}\ \bibnamefont {Glazman}}, \bibinfo {author} {\bibfnamefont {S.~M.}\
  \bibnamefont {Girvin}}, \bibinfo {author} {\bibfnamefont {M.~H.}\
  \bibnamefont {Devoret}}, \ and\ \bibinfo {author} {\bibfnamefont {R.~J.}\
  \bibnamefont {Schoelkopf}},\ }\bibfield  {title} {\enquote {\bibinfo {title}
  {{Observation of high coherence in Josephson junction qubits measured in a
  three-dimensional circuit QED architecture}},}\ }\href {\doibase
  10.1103/PhysRevLett.107.240501} {\bibfield  {journal} {\bibinfo  {journal}
  {Physical Review Letters}\ }\textbf {\bibinfo {volume} {107}},\ \bibinfo
  {pages} {240501} (\bibinfo {year} {2011})}\BibitemShut {NoStop}%
\bibitem [{\citenamefont {Barends}\ \emph {et~al.}(2013)\citenamefont
  {Barends}, \citenamefont {Kelly}, \citenamefont {Megrant}, \citenamefont
  {Sank}, \citenamefont {Jeffrey}, \citenamefont {Chen}, \citenamefont {Yin},
  \citenamefont {Chiaro}, \citenamefont {Mutus}, \citenamefont {Neill},
  \citenamefont {O'Malley}, \citenamefont {Roushan}, \citenamefont {Wenner},
  \citenamefont {White}, \citenamefont {Cleland},\ and\ \citenamefont
  {Martinis}}]{Barends2013}%
  \BibitemOpen
  \bibfield  {author} {\bibinfo {author} {\bibfnamefont {R.}~\bibnamefont
  {Barends}}, \bibinfo {author} {\bibfnamefont {J.}~\bibnamefont {Kelly}},
  \bibinfo {author} {\bibfnamefont {A.}~\bibnamefont {Megrant}}, \bibinfo
  {author} {\bibfnamefont {D.}~\bibnamefont {Sank}}, \bibinfo {author}
  {\bibfnamefont {E.}~\bibnamefont {Jeffrey}}, \bibinfo {author} {\bibfnamefont
  {Y.}~\bibnamefont {Chen}}, \bibinfo {author} {\bibfnamefont {Y.}~\bibnamefont
  {Yin}}, \bibinfo {author} {\bibfnamefont {B.}~\bibnamefont {Chiaro}},
  \bibinfo {author} {\bibfnamefont {J.}~\bibnamefont {Mutus}}, \bibinfo
  {author} {\bibfnamefont {C.}~\bibnamefont {Neill}}, \bibinfo {author}
  {\bibfnamefont {P.}~\bibnamefont {O'Malley}}, \bibinfo {author}
  {\bibfnamefont {P.}~\bibnamefont {Roushan}}, \bibinfo {author} {\bibfnamefont
  {J.}~\bibnamefont {Wenner}}, \bibinfo {author} {\bibfnamefont {T.~C.}\
  \bibnamefont {White}}, \bibinfo {author} {\bibfnamefont {A.~N.}\ \bibnamefont
  {Cleland}}, \ and\ \bibinfo {author} {\bibfnamefont {J.~M.}\ \bibnamefont
  {Martinis}},\ }\bibfield  {title} {\enquote {\bibinfo {title} {{Coherent
  josephson qubit suitable for scalable quantum integrated circuits}},}\ }\href
  {\doibase 10.1103/PhysRevLett.111.080502} {\bibfield  {journal} {\bibinfo
  {journal} {Physical Review Letters}\ }\textbf {\bibinfo {volume} {111}},\
  \bibinfo {pages} {080502} (\bibinfo {year} {2013})}\BibitemShut {NoStop}%
\bibitem [{\citenamefont {Pop}\ \emph {et~al.}(2014)\citenamefont {Pop},
  \citenamefont {Geerlings}, \citenamefont {Catelani}, \citenamefont
  {Schoelkopf}, \citenamefont {Glazman},\ and\ \citenamefont
  {Devoret}}]{Pop2014}%
  \BibitemOpen
  \bibfield  {author} {\bibinfo {author} {\bibfnamefont {I.~M.}\ \bibnamefont
  {Pop}}, \bibinfo {author} {\bibfnamefont {K.}~\bibnamefont {Geerlings}},
  \bibinfo {author} {\bibfnamefont {G.}~\bibnamefont {Catelani}}, \bibinfo
  {author} {\bibfnamefont {R.~J.}\ \bibnamefont {Schoelkopf}}, \bibinfo
  {author} {\bibfnamefont {L.~I.}\ \bibnamefont {Glazman}}, \ and\ \bibinfo
  {author} {\bibfnamefont {M.~H.}\ \bibnamefont {Devoret}},\ }\bibfield
  {title} {\enquote {\bibinfo {title} {{Coherent suppression of electromagnetic
  dissipation due to superconducting quasiparticles.}}}\ }\href {\doibase
  10.1038/nature13017} {\bibfield  {journal} {\bibinfo  {journal} {Nature}\
  }\textbf {\bibinfo {volume} {508}},\ \bibinfo {pages} {369--72} (\bibinfo
  {year} {2014})}\BibitemShut {NoStop}%
\bibitem [{\citenamefont {Yan}\ \emph {et~al.}(2015)\citenamefont {Yan},
  \citenamefont {Gustavsson}, \citenamefont {Kamal}, \citenamefont {Birenbaum},
  \citenamefont {Sears}, \citenamefont {Hover}, \citenamefont {Rosenberg},
  \citenamefont {Samach}, \citenamefont {Gudmundsen}, \citenamefont {Yoder},
  \citenamefont {Orlando}, \citenamefont {Clarke}, \citenamefont {Kerman},\
  and\ \citenamefont {Oliver}}]{Yan2015}%
  \BibitemOpen
  \bibfield  {author} {\bibinfo {author} {\bibfnamefont {F.}~\bibnamefont
  {Yan}}, \bibinfo {author} {\bibfnamefont {S.}~\bibnamefont {Gustavsson}},
  \bibinfo {author} {\bibfnamefont {A.}~\bibnamefont {Kamal}}, \bibinfo
  {author} {\bibfnamefont {J.}~\bibnamefont {Birenbaum}}, \bibinfo {author}
  {\bibfnamefont {A.~P.}\ \bibnamefont {Sears}}, \bibinfo {author}
  {\bibfnamefont {D.}~\bibnamefont {Hover}}, \bibinfo {author} {\bibfnamefont
  {D.}~\bibnamefont {Rosenberg}}, \bibinfo {author} {\bibfnamefont
  {G.}~\bibnamefont {Samach}}, \bibinfo {author} {\bibfnamefont {T.~J.}\
  \bibnamefont {Gudmundsen}}, \bibinfo {author} {\bibfnamefont {J.~L.}\
  \bibnamefont {Yoder}}, \bibinfo {author} {\bibfnamefont {T.~P.}\ \bibnamefont
  {Orlando}}, \bibinfo {author} {\bibfnamefont {J.}~\bibnamefont {Clarke}},
  \bibinfo {author} {\bibfnamefont {A.~J.}\ \bibnamefont {Kerman}}, \ and\
  \bibinfo {author} {\bibfnamefont {W.~D.}\ \bibnamefont {Oliver}},\ }\bibfield
   {title} {\enquote {\bibinfo {title} {{The Flux Qubit Revisited to Enhance
  Coherence and Reproducibility}},}\ }\href {\doibase 10.1038/ncomms12964}
  {\bibfield  {journal} {\bibinfo  {journal} {Nature Communications}\ }\textbf
  {\bibinfo {volume} {7}},\ \bibinfo {pages} {12964} (\bibinfo {year}
  {2015})}\BibitemShut {NoStop}%
\bibitem [{\citenamefont {Foroozani}\ \emph {et~al.}(2019)\citenamefont
  {Foroozani}, \citenamefont {Hobbs}, \citenamefont {Hung}, \citenamefont
  {Olson}, \citenamefont {Ashworth}, \citenamefont {Holland}, \citenamefont
  {Malloy}, \citenamefont {Kearney}, \citenamefont {O'Brien}, \citenamefont
  {Bunday}, \citenamefont {DiPaola}, \citenamefont {Advocate}, \citenamefont
  {Murray}, \citenamefont {Hansen}, \citenamefont {Novak}, \citenamefont
  {Bennett}, \citenamefont {Rodgers}, \citenamefont {Baker-O'Neal},
  \citenamefont {Sapp}, \citenamefont {Barth}, \citenamefont {Hedrick},
  \citenamefont {Goldblatt}, \citenamefont {Rao},\ and\ \citenamefont
  {Osborn}}]{Foroozani2019}%
  \BibitemOpen
  \bibfield  {author} {\bibinfo {author} {\bibfnamefont {N.}~\bibnamefont
  {Foroozani}}, \bibinfo {author} {\bibfnamefont {C.}~\bibnamefont {Hobbs}},
  \bibinfo {author} {\bibfnamefont {C.~C.}\ \bibnamefont {Hung}}, \bibinfo
  {author} {\bibfnamefont {S.}~\bibnamefont {Olson}}, \bibinfo {author}
  {\bibfnamefont {D.}~\bibnamefont {Ashworth}}, \bibinfo {author}
  {\bibfnamefont {E.}~\bibnamefont {Holland}}, \bibinfo {author} {\bibfnamefont
  {M.}~\bibnamefont {Malloy}}, \bibinfo {author} {\bibfnamefont
  {P.}~\bibnamefont {Kearney}}, \bibinfo {author} {\bibfnamefont
  {B.}~\bibnamefont {O'Brien}}, \bibinfo {author} {\bibfnamefont
  {B.}~\bibnamefont {Bunday}}, \bibinfo {author} {\bibfnamefont
  {D.}~\bibnamefont {DiPaola}}, \bibinfo {author} {\bibfnamefont
  {W.}~\bibnamefont {Advocate}}, \bibinfo {author} {\bibfnamefont
  {T.}~\bibnamefont {Murray}}, \bibinfo {author} {\bibfnamefont
  {P.}~\bibnamefont {Hansen}}, \bibinfo {author} {\bibfnamefont
  {S.}~\bibnamefont {Novak}}, \bibinfo {author} {\bibfnamefont
  {S.}~\bibnamefont {Bennett}}, \bibinfo {author} {\bibfnamefont
  {M.}~\bibnamefont {Rodgers}}, \bibinfo {author} {\bibfnamefont
  {B.}~\bibnamefont {Baker-O'Neal}}, \bibinfo {author} {\bibfnamefont
  {B.}~\bibnamefont {Sapp}}, \bibinfo {author} {\bibfnamefont {E.}~\bibnamefont
  {Barth}}, \bibinfo {author} {\bibfnamefont {J.}~\bibnamefont {Hedrick}},
  \bibinfo {author} {\bibfnamefont {R.}~\bibnamefont {Goldblatt}}, \bibinfo
  {author} {\bibfnamefont {S.~S.~P.}\ \bibnamefont {Rao}}, \ and\ \bibinfo
  {author} {\bibfnamefont {K.~D.}\ \bibnamefont {Osborn}},\ }\bibfield  {title}
  {\enquote {\bibinfo {title} {{Development of transmon qubits solely from
  optical lithography on 300 mm wafers}},}\ }\href {\doibase
  10.1088/2058-9565/ab0ca8} {\bibfield  {journal} {\bibinfo  {journal} {Quantum
  Science and Technology}\ }\textbf {\bibinfo {volume} {4}},\ \bibinfo {pages}
  {025012} (\bibinfo {year} {2019})}\BibitemShut {NoStop}%
\bibitem [{\citenamefont {Vazquez-Mena}\ \emph {et~al.}(2015)\citenamefont
  {Vazquez-Mena}, \citenamefont {Gross}, \citenamefont {Xie}, \citenamefont
  {Villanueva},\ and\ \citenamefont {Brugger}}]{Vazquez2015}%
  \BibitemOpen
  \bibfield  {author} {\bibinfo {author} {\bibfnamefont {O.}~\bibnamefont
  {Vazquez-Mena}}, \bibinfo {author} {\bibfnamefont {L.}~\bibnamefont {Gross}},
  \bibinfo {author} {\bibfnamefont {S.}~\bibnamefont {Xie}}, \bibinfo {author}
  {\bibfnamefont {L.}~\bibnamefont {Villanueva}}, \ and\ \bibinfo {author}
  {\bibfnamefont {J.}~\bibnamefont {Brugger}},\ }\bibfield  {title} {\enquote
  {\bibinfo {title} {{Resistless nanofabrication by stencil lithography: A
  review}},}\ }\href {\doibase 10.1016/j.mee.2014.08.003} {\bibfield  {journal}
  {\bibinfo  {journal} {Microelectronic Engineering}\ }\textbf {\bibinfo
  {volume} {132}},\ \bibinfo {pages} {236--254} (\bibinfo {year}
  {2015})}\BibitemShut {NoStop}%
\bibitem [{\citenamefont {Place}\ \emph {et~al.}(2020)\citenamefont {Place},
  \citenamefont {Rodgers}, \citenamefont {Mundada}, \citenamefont {Smitham},
  \citenamefont {Fitzpatrick}, \citenamefont {Leng}, \citenamefont {Premkumar},
  \citenamefont {Bryon}, \citenamefont {Sussman}, \citenamefont {Cheng},
  \citenamefont {Madhavan}, \citenamefont {Babla}, \citenamefont {Jaeck},
  \citenamefont {Gyenis}, \citenamefont {Yao}, \citenamefont {Cava},
  \citenamefont {de~Leon},\ and\ \citenamefont {Houck}}]{Place2020}%
  \BibitemOpen
  \bibfield  {author} {\bibinfo {author} {\bibfnamefont {A.~P.~M.}\
  \bibnamefont {Place}}, \bibinfo {author} {\bibfnamefont {L.~V.~H.}\
  \bibnamefont {Rodgers}}, \bibinfo {author} {\bibfnamefont {P.}~\bibnamefont
  {Mundada}}, \bibinfo {author} {\bibfnamefont {B.~M.}\ \bibnamefont
  {Smitham}}, \bibinfo {author} {\bibfnamefont {M.}~\bibnamefont
  {Fitzpatrick}}, \bibinfo {author} {\bibfnamefont {Z.}~\bibnamefont {Leng}},
  \bibinfo {author} {\bibfnamefont {A.}~\bibnamefont {Premkumar}}, \bibinfo
  {author} {\bibfnamefont {J.}~\bibnamefont {Bryon}}, \bibinfo {author}
  {\bibfnamefont {S.}~\bibnamefont {Sussman}}, \bibinfo {author} {\bibfnamefont
  {G.}~\bibnamefont {Cheng}}, \bibinfo {author} {\bibfnamefont
  {T.}~\bibnamefont {Madhavan}}, \bibinfo {author} {\bibfnamefont {H.~K.}\
  \bibnamefont {Babla}}, \bibinfo {author} {\bibfnamefont {B.}~\bibnamefont
  {Jaeck}}, \bibinfo {author} {\bibfnamefont {A.}~\bibnamefont {Gyenis}},
  \bibinfo {author} {\bibfnamefont {N.}~\bibnamefont {Yao}}, \bibinfo {author}
  {\bibfnamefont {R.~J.}\ \bibnamefont {Cava}}, \bibinfo {author}
  {\bibfnamefont {N.~P.}\ \bibnamefont {de~Leon}}, \ and\ \bibinfo {author}
  {\bibfnamefont {A.~A.}\ \bibnamefont {Houck}},\ }\bibfield  {title} {\enquote
  {\bibinfo {title} {{New material platform for superconducting transmon qubits
  with coherence times exceeding 0.3 milliseconds}},}\ }\href
  {http://arxiv.org/abs/2003.00024} {\  (\bibinfo {year} {2020})},\ \Eprint
  {http://arxiv.org/abs/2003.00024} {arXiv:2003.00024} \BibitemShut {NoStop}%
\bibitem [{Nor()}]{Norcada}%
  \BibitemOpen
  \href@noop {} {\enquote {\bibinfo {title} {{Norcada Inc., Address: 4548-99
  Street, Edmonton, AB T6E 5H5, Canada Email: info@norcada.com, Web:
  www.norcada.com}},}\ }\BibitemShut {NoStop}%
\bibitem [{\citenamefont {Laermer}\ and\ \citenamefont {Schilp}(1996)}]{Bosch}%
  \BibitemOpen
  \bibfield  {author} {\bibinfo {author} {\bibfnamefont {F.}~\bibnamefont
  {Laermer}}\ and\ \bibinfo {author} {\bibfnamefont {A.}~\bibnamefont
  {Schilp}},\ }\bibfield  {title} {\enquote {\bibinfo {title} {{Method of
  anisotropically etching silicon}},}\ }\href@noop {} {\bibfield  {journal}
  {\bibinfo  {journal} {U.S. patent US5501893A}\ } (\bibinfo {year}
  {1996})}\BibitemShut {NoStop}%
\bibitem [{\citenamefont {Gladchenko}\ \emph {et~al.}(2009)\citenamefont
  {Gladchenko}, \citenamefont {Olaya}, \citenamefont {Dupont-Ferrier},
  \citenamefont {Dou{\c{c}}ot}, \citenamefont {Ioffe},\ and\ \citenamefont
  {Gershenson}}]{Gladchenko2009}%
  \BibitemOpen
  \bibfield  {author} {\bibinfo {author} {\bibfnamefont {S.}~\bibnamefont
  {Gladchenko}}, \bibinfo {author} {\bibfnamefont {D.}~\bibnamefont {Olaya}},
  \bibinfo {author} {\bibfnamefont {E.}~\bibnamefont {Dupont-Ferrier}},
  \bibinfo {author} {\bibfnamefont {B.}~\bibnamefont {Dou{\c{c}}ot}}, \bibinfo
  {author} {\bibfnamefont {L.~B.}\ \bibnamefont {Ioffe}}, \ and\ \bibinfo
  {author} {\bibfnamefont {M.~E.}\ \bibnamefont {Gershenson}},\ }\bibfield
  {title} {\enquote {\bibinfo {title} {{Superconducting nanocircuits for
  topologically protected qubits}},}\ }\href {\doibase 10.1038/nphys1151}
  {\bibfield  {journal} {\bibinfo  {journal} {Nature Physics}\ }\textbf
  {\bibinfo {volume} {5}},\ \bibinfo {pages} {48--53} (\bibinfo {year}
  {2009})}\BibitemShut {NoStop}%
\bibitem [{\citenamefont {Yeom}\ \emph {et~al.}(2006)\citenamefont {Yeom},
  \citenamefont {Wu}, \citenamefont {Selby},\ and\ \citenamefont
  {Shannon}}]{Yeom2005}%
  \BibitemOpen
  \bibfield  {author} {\bibinfo {author} {\bibfnamefont {J.}~\bibnamefont
  {Yeom}}, \bibinfo {author} {\bibfnamefont {Y.}~\bibnamefont {Wu}}, \bibinfo
  {author} {\bibfnamefont {J.~C.}\ \bibnamefont {Selby}}, \ and\ \bibinfo
  {author} {\bibfnamefont {M.~A.}\ \bibnamefont {Shannon}},\ }\bibfield
  {title} {\enquote {\bibinfo {title} {{Maximum achievable aspect ratio in deep
  reactive ion etching of silicon due to aspect ratio dependent transport and
  the microloading effect}},}\ }\href {\doibase 10.1116/1.2101678} {\bibfield
  {journal} {\bibinfo  {journal} {Journal of Vacuum Science {\&} Technology B:
  Microelectronics and Nanometer Structures}\ }\textbf {\bibinfo {volume}
  {23}},\ \bibinfo {pages} {2319} (\bibinfo {year} {2006})}\BibitemShut
  {NoStop}%
\bibitem [{\citenamefont {Iida}\ \emph {et~al.}(2000)\citenamefont {Iida},
  \citenamefont {Itoh}, \citenamefont {Noguchi},\ and\ \citenamefont
  {Takano}}]{Tsutomu2000}%
  \BibitemOpen
  \bibfield  {author} {\bibinfo {author} {\bibfnamefont {T.}~\bibnamefont
  {Iida}}, \bibinfo {author} {\bibfnamefont {T.}~\bibnamefont {Itoh}}, \bibinfo
  {author} {\bibfnamefont {D.}~\bibnamefont {Noguchi}}, \ and\ \bibinfo
  {author} {\bibfnamefont {Y.}~\bibnamefont {Takano}},\ }\bibfield  {title}
  {\enquote {\bibinfo {title} {{Residual lattice strain in thin
  silicon-on-insulator bonded wafers: Thermal behavior and formation
  mechanisms}},}\ }\href {\doibase 10.1063/1.371925} {\bibfield  {journal}
  {\bibinfo  {journal} {Journal of Applied Physics}\ }\textbf {\bibinfo
  {volume} {87}},\ \bibinfo {pages} {675--681} (\bibinfo {year}
  {2000})}\BibitemShut {NoStop}%
\bibitem [{\citenamefont {Park}\ \emph {et~al.}(2010)\citenamefont {Park},
  \citenamefont {Brugger}, \citenamefont {{Guillermo Villanueva}},
  \citenamefont {Savu}, \citenamefont {Sidler},\ and\ \citenamefont
  {Vazquez-Mena}}]{Park2010}%
  \BibitemOpen
  \bibfield  {author} {\bibinfo {author} {\bibfnamefont {C.~W.}\ \bibnamefont
  {Park}}, \bibinfo {author} {\bibfnamefont {J.}~\bibnamefont {Brugger}},
  \bibinfo {author} {\bibfnamefont {L.}~\bibnamefont {{Guillermo Villanueva}}},
  \bibinfo {author} {\bibfnamefont {V.}~\bibnamefont {Savu}}, \bibinfo {author}
  {\bibfnamefont {K.}~\bibnamefont {Sidler}}, \ and\ \bibinfo {author}
  {\bibfnamefont {O.}~\bibnamefont {Vazquez-Mena}},\ }\bibfield  {title}
  {\enquote {\bibinfo {title} {{Reliable and Improved Nanoscale Stencil
  Lithography by Membrane Stabilization, Blurring, and Clogging
  Corrections}},}\ }\href {\doibase 10.1109/tnano.2010.2042724} {\bibfield
  {journal} {\bibinfo  {journal} {IEEE Transactions on Nanotechnology}\
  }\textbf {\bibinfo {volume} {10}},\ \bibinfo {pages} {352--357} (\bibinfo
  {year} {2010})}\BibitemShut {NoStop}%
\bibitem [{\citenamefont {Blais}\ \emph {et~al.}(2004)\citenamefont {Blais},
  \citenamefont {Huang}, \citenamefont {Wallraff}, \citenamefont {Girvin},\
  and\ \citenamefont {Schoelkopf}}]{Blais2004}%
  \BibitemOpen
  \bibfield  {author} {\bibinfo {author} {\bibfnamefont {A.}~\bibnamefont
  {Blais}}, \bibinfo {author} {\bibfnamefont {R.~S.}\ \bibnamefont {Huang}},
  \bibinfo {author} {\bibfnamefont {A.}~\bibnamefont {Wallraff}}, \bibinfo
  {author} {\bibfnamefont {S.~M.}\ \bibnamefont {Girvin}}, \ and\ \bibinfo
  {author} {\bibfnamefont {R.~J.}\ \bibnamefont {Schoelkopf}},\ }\bibfield
  {title} {\enquote {\bibinfo {title} {{Cavity quantum electrodynamics for
  superconducting electrical circuits: An architecture for quantum
  computation}},}\ }\href {\doibase 10.1103/PhysRevA.69.062320} {\bibfield
  {journal} {\bibinfo  {journal} {Physical Review A}\ }\textbf {\bibinfo
  {volume} {69}},\ \bibinfo {pages} {062320} (\bibinfo {year}
  {2004})}\BibitemShut {NoStop}%
\bibitem [{\citenamefont {Sivak}\ \emph {et~al.}(2020)\citenamefont {Sivak},
  \citenamefont {Shankar}, \citenamefont {Liu}, \citenamefont {Aumentado},\
  and\ \citenamefont {Devoret}}]{Sivak2020}%
  \BibitemOpen
  \bibfield  {author} {\bibinfo {author} {\bibfnamefont {V.~V.}\ \bibnamefont
  {Sivak}}, \bibinfo {author} {\bibfnamefont {S.}~\bibnamefont {Shankar}},
  \bibinfo {author} {\bibfnamefont {G.}~\bibnamefont {Liu}}, \bibinfo {author}
  {\bibfnamefont {J.}~\bibnamefont {Aumentado}}, \ and\ \bibinfo {author}
  {\bibfnamefont {M.~H.}\ \bibnamefont {Devoret}},\ }\bibfield  {title}
  {\enquote {\bibinfo {title} {{Josephson Array-Mode Parametric Amplifier}},}\
  }\href {\doibase 10.1103/PhysRevApplied.13.024014} {\bibfield  {journal}
  {\bibinfo  {journal} {Physical Review Applied}\ }\textbf {\bibinfo {volume}
  {13}},\ \bibinfo {pages} {024014} (\bibinfo {year} {2020})}\BibitemShut
  {NoStop}%
\bibitem [{\citenamefont {Serniak}\ \emph {et~al.}(2019)\citenamefont
  {Serniak}, \citenamefont {Diamond}, \citenamefont {Hays}, \citenamefont
  {Fatemi}, \citenamefont {Shankar}, \citenamefont {Frunzio}, \citenamefont
  {Schoelkopf},\ and\ \citenamefont {Devoret}}]{Kyle2019}%
  \BibitemOpen
  \bibfield  {author} {\bibinfo {author} {\bibfnamefont {K.}~\bibnamefont
  {Serniak}}, \bibinfo {author} {\bibfnamefont {S.}~\bibnamefont {Diamond}},
  \bibinfo {author} {\bibfnamefont {M.}~\bibnamefont {Hays}}, \bibinfo {author}
  {\bibfnamefont {V.}~\bibnamefont {Fatemi}}, \bibinfo {author} {\bibfnamefont
  {S.}~\bibnamefont {Shankar}}, \bibinfo {author} {\bibfnamefont
  {L.}~\bibnamefont {Frunzio}}, \bibinfo {author} {\bibfnamefont
  {R.}~\bibnamefont {Schoelkopf}}, \ and\ \bibinfo {author} {\bibfnamefont
  {M.}~\bibnamefont {Devoret}},\ }\bibfield  {title} {\enquote {\bibinfo
  {title} {{Direct Dispersive Monitoring of Charge Parity in
  Offset-Charge-Sensitive Transmons}},}\ }\href {\doibase
  10.1103/PhysRevApplied.12.014052} {\bibfield  {journal} {\bibinfo  {journal}
  {Physical Review Applied}\ }\textbf {\bibinfo {volume} {12}},\ \bibinfo
  {pages} {014052} (\bibinfo {year} {2019})}\BibitemShut {NoStop}%
\bibitem [{\citenamefont {Serniak}\ \emph {et~al.}(2018)\citenamefont
  {Serniak}, \citenamefont {Hays}, \citenamefont {{De Lange}}, \citenamefont
  {Diamond}, \citenamefont {Shankar}, \citenamefont {Burkhart}, \citenamefont
  {Frunzio}, \citenamefont {Houzet},\ and\ \citenamefont {Devoret}}]{Kyle2018}%
  \BibitemOpen
  \bibfield  {author} {\bibinfo {author} {\bibfnamefont {K.}~\bibnamefont
  {Serniak}}, \bibinfo {author} {\bibfnamefont {M.}~\bibnamefont {Hays}},
  \bibinfo {author} {\bibfnamefont {G.}~\bibnamefont {{De Lange}}}, \bibinfo
  {author} {\bibfnamefont {S.}~\bibnamefont {Diamond}}, \bibinfo {author}
  {\bibfnamefont {S.}~\bibnamefont {Shankar}}, \bibinfo {author} {\bibfnamefont
  {L.~D.}\ \bibnamefont {Burkhart}}, \bibinfo {author} {\bibfnamefont
  {L.}~\bibnamefont {Frunzio}}, \bibinfo {author} {\bibfnamefont
  {M.}~\bibnamefont {Houzet}}, \ and\ \bibinfo {author} {\bibfnamefont {M.~H.}\
  \bibnamefont {Devoret}},\ }\bibfield  {title} {\enquote {\bibinfo {title}
  {{Hot Nonequilibrium Quasiparticles in Transmon Qubits}},}\ }\href {\doibase
  10.1103/PhysRevLett.121.157701} {\bibfield  {journal} {\bibinfo  {journal}
  {Physical Review Letters}\ }\textbf {\bibinfo {volume} {121}},\ \bibinfo
  {pages} {157701} (\bibinfo {year} {2018})}\BibitemShut {NoStop}%
\bibitem [{\citenamefont {Klimov}\ \emph {et~al.}(2018)\citenamefont {Klimov},
  \citenamefont {Kelly}, \citenamefont {Chen}, \citenamefont {Neeley},
  \citenamefont {Megrant}, \citenamefont {Burkett}, \citenamefont {Barends},
  \citenamefont {Arya}, \citenamefont {Chiaro}, \citenamefont {Chen},
  \citenamefont {Dunsworth}, \citenamefont {Fowler}, \citenamefont {Foxen},
  \citenamefont {Gidney}, \citenamefont {Giustina}, \citenamefont {Graff},
  \citenamefont {Huang}, \citenamefont {Jeffrey}, \citenamefont {Lucero},
  \citenamefont {Mutus}, \citenamefont {Naaman}, \citenamefont {Neill},
  \citenamefont {Quintana}, \citenamefont {Roushan}, \citenamefont {Sank},
  \citenamefont {Vainsencher}, \citenamefont {Wenner}, \citenamefont {White},
  \citenamefont {Boixo}, \citenamefont {Babbush}, \citenamefont {Smelyanskiy},
  \citenamefont {Neven},\ and\ \citenamefont {Martinis}}]{Klimov2018}%
  \BibitemOpen
  \bibfield  {author} {\bibinfo {author} {\bibfnamefont {P.~V.}\ \bibnamefont
  {Klimov}}, \bibinfo {author} {\bibfnamefont {J.}~\bibnamefont {Kelly}},
  \bibinfo {author} {\bibfnamefont {Z.}~\bibnamefont {Chen}}, \bibinfo {author}
  {\bibfnamefont {M.}~\bibnamefont {Neeley}}, \bibinfo {author} {\bibfnamefont
  {A.}~\bibnamefont {Megrant}}, \bibinfo {author} {\bibfnamefont
  {B.}~\bibnamefont {Burkett}}, \bibinfo {author} {\bibfnamefont
  {R.}~\bibnamefont {Barends}}, \bibinfo {author} {\bibfnamefont
  {K.}~\bibnamefont {Arya}}, \bibinfo {author} {\bibfnamefont {B.}~\bibnamefont
  {Chiaro}}, \bibinfo {author} {\bibfnamefont {Y.}~\bibnamefont {Chen}},
  \bibinfo {author} {\bibfnamefont {A.}~\bibnamefont {Dunsworth}}, \bibinfo
  {author} {\bibfnamefont {A.}~\bibnamefont {Fowler}}, \bibinfo {author}
  {\bibfnamefont {B.}~\bibnamefont {Foxen}}, \bibinfo {author} {\bibfnamefont
  {C.}~\bibnamefont {Gidney}}, \bibinfo {author} {\bibfnamefont
  {M.}~\bibnamefont {Giustina}}, \bibinfo {author} {\bibfnamefont
  {R.}~\bibnamefont {Graff}}, \bibinfo {author} {\bibfnamefont
  {T.}~\bibnamefont {Huang}}, \bibinfo {author} {\bibfnamefont
  {E.}~\bibnamefont {Jeffrey}}, \bibinfo {author} {\bibfnamefont
  {E.}~\bibnamefont {Lucero}}, \bibinfo {author} {\bibfnamefont {J.~Y.}\
  \bibnamefont {Mutus}}, \bibinfo {author} {\bibfnamefont {O.}~\bibnamefont
  {Naaman}}, \bibinfo {author} {\bibfnamefont {C.}~\bibnamefont {Neill}},
  \bibinfo {author} {\bibfnamefont {C.}~\bibnamefont {Quintana}}, \bibinfo
  {author} {\bibfnamefont {P.}~\bibnamefont {Roushan}}, \bibinfo {author}
  {\bibfnamefont {D.}~\bibnamefont {Sank}}, \bibinfo {author} {\bibfnamefont
  {A.}~\bibnamefont {Vainsencher}}, \bibinfo {author} {\bibfnamefont
  {J.}~\bibnamefont {Wenner}}, \bibinfo {author} {\bibfnamefont {T.~C.}\
  \bibnamefont {White}}, \bibinfo {author} {\bibfnamefont {S.}~\bibnamefont
  {Boixo}}, \bibinfo {author} {\bibfnamefont {R.}~\bibnamefont {Babbush}},
  \bibinfo {author} {\bibfnamefont {V.~N.}\ \bibnamefont {Smelyanskiy}},
  \bibinfo {author} {\bibfnamefont {H.}~\bibnamefont {Neven}}, \ and\ \bibinfo
  {author} {\bibfnamefont {J.~M.}\ \bibnamefont {Martinis}},\ }\bibfield
  {title} {\enquote {\bibinfo {title} {{Fluctuations of Energy-Relaxation Times
  in Superconducting Qubits}},}\ }\href {\doibase
  10.1103/PhysRevLett.121.090502} {\bibfield  {journal} {\bibinfo  {journal}
  {Physical Review Letters}\ }\textbf {\bibinfo {volume} {121}},\ \bibinfo
  {pages} {090502} (\bibinfo {year} {2018})}\BibitemShut {NoStop}%
\bibitem [{\citenamefont {Wang}\ \emph {et~al.}(2019)\citenamefont {Wang},
  \citenamefont {Shankar}, \citenamefont {Minev}, \citenamefont
  {Campagne-Ibarcq}, \citenamefont {Narla},\ and\ \citenamefont
  {Devoret}}]{Zhixin2018}%
  \BibitemOpen
  \bibfield  {author} {\bibinfo {author} {\bibfnamefont {Z.}~\bibnamefont
  {Wang}}, \bibinfo {author} {\bibfnamefont {S.}~\bibnamefont {Shankar}},
  \bibinfo {author} {\bibfnamefont {Z.~K.}\ \bibnamefont {Minev}}, \bibinfo
  {author} {\bibfnamefont {P.}~\bibnamefont {Campagne-Ibarcq}}, \bibinfo
  {author} {\bibfnamefont {A.}~\bibnamefont {Narla}}, \ and\ \bibinfo {author}
  {\bibfnamefont {M.~H.}\ \bibnamefont {Devoret}},\ }\bibfield  {title}
  {\enquote {\bibinfo {title} {{Cavity Attenuators for Superconducting
  Qubits}},}\ }\href {\doibase 10.1103/PhysRevApplied.11.014031} {\bibfield
  {journal} {\bibinfo  {journal} {Physical Review Applied}\ }\textbf {\bibinfo
  {volume} {11}},\ \bibinfo {pages} {014031} (\bibinfo {year}
  {2019})}\BibitemShut {NoStop}%
\bibitem [{\citenamefont {Ootuka}\ \emph {et~al.}(1996)\citenamefont {Ootuka},
  \citenamefont {Ono}, \citenamefont {Shimada},\ and\ \citenamefont
  {Kobayashi}}]{Ootuka1996}%
  \BibitemOpen
  \bibfield  {author} {\bibinfo {author} {\bibfnamefont {Y.}~\bibnamefont
  {Ootuka}}, \bibinfo {author} {\bibfnamefont {K.}~\bibnamefont {Ono}},
  \bibinfo {author} {\bibfnamefont {H.}~\bibnamefont {Shimada}}, \ and\
  \bibinfo {author} {\bibfnamefont {S.-I.}\ \bibnamefont {Kobayashi}},\
  }\bibfield  {title} {\enquote {\bibinfo {title} {{A new fabrication method of
  ultra small tunnel junctions}},}\ }\href {\doibase
  10.1016/0921-4526(96)00427-9} {\bibfield  {journal} {\bibinfo  {journal}
  {Physica B}\ }\textbf {\bibinfo {volume} {227}},\ \bibinfo {pages} {307--309}
  (\bibinfo {year} {1996})}\BibitemShut {NoStop}%
\bibitem [{\citenamefont {Savu}\ \emph {et~al.}(2009)\citenamefont {Savu},
  \citenamefont {Kivioja}, \citenamefont {Ahopelto},\ and\ \citenamefont
  {Brugger}}]{Savu2009}%
  \BibitemOpen
  \bibfield  {author} {\bibinfo {author} {\bibfnamefont {V.}~\bibnamefont
  {Savu}}, \bibinfo {author} {\bibfnamefont {J.}~\bibnamefont {Kivioja}},
  \bibinfo {author} {\bibfnamefont {J.}~\bibnamefont {Ahopelto}}, \ and\
  \bibinfo {author} {\bibfnamefont {J.}~\bibnamefont {Brugger}},\ }\bibfield
  {title} {\enquote {\bibinfo {title} {{Quick and clean: stencil lithography
  for wafer-scale fabrication of superconducting tunnel junctions}},}\ }\href
  {\doibase 10.1109/TASC.2009.2019075} {\bibfield  {journal} {\bibinfo
  {journal} {IEEE Transactions on Applied Superconductivity}\ }\textbf
  {\bibinfo {volume} {19}},\ \bibinfo {pages} {242--244} (\bibinfo {year}
  {2009})}\BibitemShut {NoStop}%
\bibitem [{\citenamefont {Welander}\ \emph {et~al.}(2012)\citenamefont
  {Welander}, \citenamefont {Bolkhovsky}, \citenamefont {Weir}, \citenamefont
  {Gouker},\ and\ \citenamefont {Oliver}}]{Welander2012}%
  \BibitemOpen
  \bibfield  {author} {\bibinfo {author} {\bibfnamefont {P.~B.}\ \bibnamefont
  {Welander}}, \bibinfo {author} {\bibfnamefont {V.}~\bibnamefont
  {Bolkhovsky}}, \bibinfo {author} {\bibfnamefont {T.~J.}\ \bibnamefont
  {Weir}}, \bibinfo {author} {\bibfnamefont {M.~A.}\ \bibnamefont {Gouker}}, \
  and\ \bibinfo {author} {\bibfnamefont {W.~D.}\ \bibnamefont {Oliver}},\
  }\bibfield  {title} {\enquote {\bibinfo {title} {{Shadow evaporation of
  epitaxial Al/Al2O3/Al tunnel junctions on sapphire utilizing an inorganic
  bilayer mask}},}\ }\href@noop {} {\  (\bibinfo {year} {2012})},\ \Eprint
  {http://arxiv.org/abs/1203.6007v1} {1203.6007v1} \BibitemShut {NoStop}%
\end{thebibliography}%

\pagebreak
\onecolumngrid
\section*{Supplementary Material}
\beginsupplement


\subsection{Experimental setup}
\begin{figure}[H]
	\centering
	\includegraphics{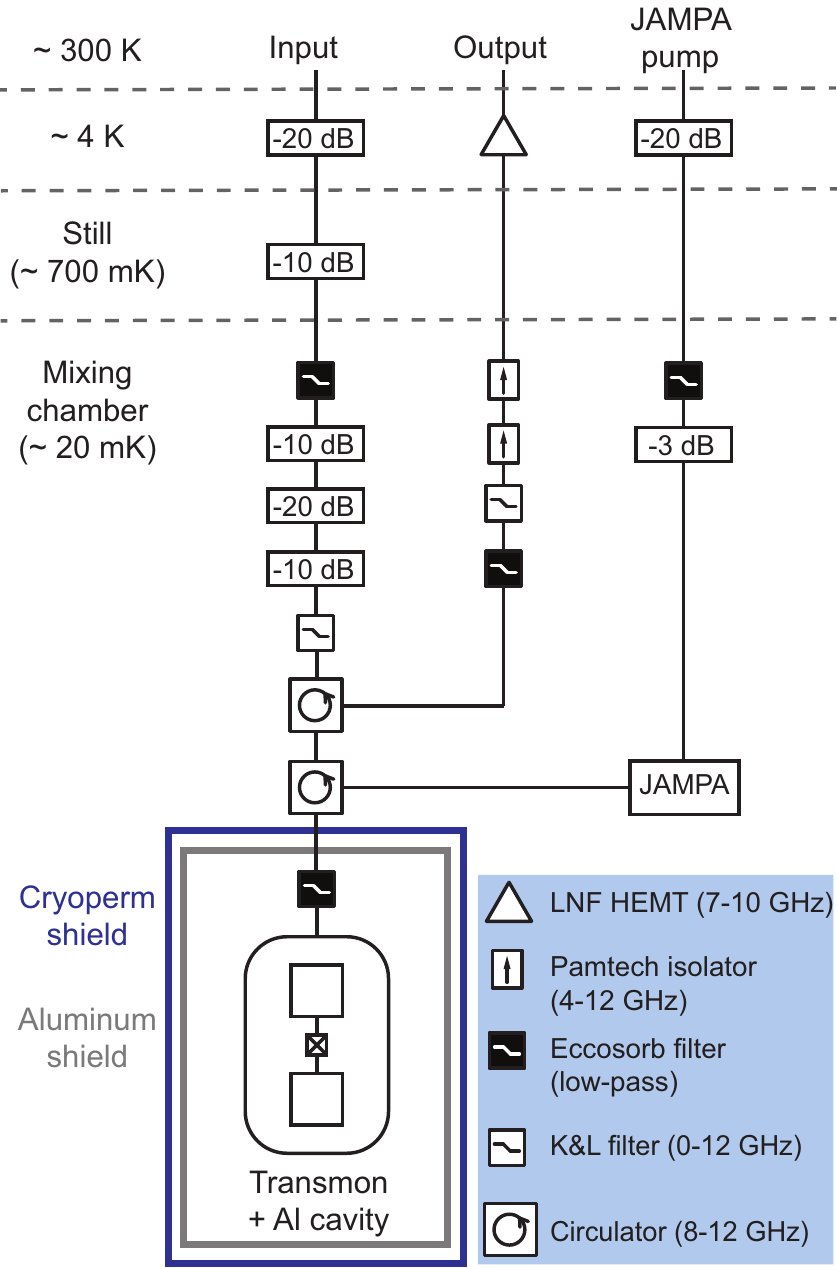}
	\caption{\label{figure_1_supp} The wiring diagram of the cryogenic microwave measurement setup. Acronyms labeling components include the Josephson Array-Mode Parametric Amplifier (JAMPA), and Low Noise Factory (LNF) High Electron Mobility Transistor (HEMT) amplifier.}
\end{figure}

The transmon chip is mounted in a 3D aluminum cavity and all together thermalized to the mixing chamber of a cryogen-free dilution refrigerator with a base temperature of approximately 20 mK (Fig.~\ref{figure_1_supp}). The fundamental mode of the 3D readout cavity is at $\omega_\mathrm{r}/2\pi = \mathrm{9.1~GHz}$ with external coupling rate set at $\kappa/2\pi = \mathrm{2.5~MHz}$. The transmon has ground-to-first-excited-state transition frequency $\omega_\mathrm{ge}/2\pi=6.01~\mathrm{GHz}$, anharmonicity  $\alpha/2\pi=\omega_\mathrm{ge} - \omega_\mathrm{ef} = 0.23~\mathrm{GHz}$, and cross-Kerr to the readout cavity mode $\chi_\mathrm{qr} = 1.2~\mathrm{MHz}$.

All the microwave pulses used to probe and control the system are routed through the input port and propagate down a heavily attenuated line. The reflected off from the 3D readout cavity signal is amplified by reflection of a near quantum limited Josephson array-mode parametric amplifier (JAMPA)\cite{Sivak2020}. Two circulators are routing the reflected signal towards the output line, where it is further amplified with a high electron mobility transistor amplifier (HEMT), before reaching the room-temperature output port, where it is detected and further processed with room-temperature electronic equipment.

\subsection{Characterization of additional devices}
To assess the reproducibility of the fabrication technique, we performed additional coherence measurement in 5 more transmons (B-F), fabricated with the same mask on two separate sapphire wafers. These measurements were performed in a different dilution refrigerator with a similar wiring as the one shown in Fig.~\ref{figure_1_supp}. However, the input line had 10 dB less attenuation at base temperature, the cavity/transmon system did not include aluminum shielding, and the signal reflected from the cavity was routed directly to the output line without the use of a quantum limited parametric amplifier. We measured $T_\mathrm{1}$, $T_\mathrm{2R}$, and $T_\mathrm{2e}$ times at least three times for each device, all yielded transmons with energy relaxation times of more than $16~\upmu \mathrm{s}$. The length of each measurement was between 3 to 12 min, depending on the number of points and averaged trajectories. Owing to the reduced signal-to-noise ratio of the readout chain in the absence of a near quantum limited parametric amplifier, the measurement lengths for transmons B-F were much longer compared to what was required for transmon A. This limitation prevented us from accumulating statistics of the coherence of transmons B-F as a function of time, as we accomplished for transmon A (Fig. 6). A summary with the full properties of all the measured transmons that have been fabricated with our new technique is given in the Table~\ref{table}. Considering the differences in the experimental apparatuses, which are described above, the coherence values of transmons B-F are not directly comparable to the ones measured for transmon A. However, the dispersion in coherence values among different devices is comparable to what we have observed in transmon devices fabricated with standard techniques. Nonetheless, further investigation is required to determine whether this behavior is related to variations in the intrinsic properties of the devices produced by the presented fabrication technique or due to the differences between the experimental apparatuses.
\begin{table}[H]
	\caption{\label{table} Measured transmon qubit parameters. From left to right, qubit transition frequency $\omega_\mathrm{ge}$, anharmonicity $\alpha=\omega_\mathrm{ge} - \omega_\mathrm{ef}$, cross-Kerr beetwen the qubit and the readout cavity mode $\chi_\mathrm{qr}$, mean energy relaxation times $\overline T_\mathrm{1}$, $\overline T_\mathrm{2R}$ Ramsey and $\overline T_\mathrm{2e}$ Hahn echo dephasing times. The coherence times of transmon D fluctuated in time significantly, therefore the range of values is presented instead of mean values.}
	\begin{ruledtabular}
		\begin{tabular}{cccccccc}
			&\shortstack{$\omega_\mathrm{ge}/\mathrm{2\pi}$ \\ (GHz)}
			&\shortstack{$\alpha/\mathrm{2\pi}$ \\ (GHz)}
			&\shortstack{$\chi/\mathrm{2\pi}$ \\ (MHz)}
			&\shortstack{$\overline T_{\mathrm{1}}$ \\ $\mathrm{(\upmu s)}$}
			&\shortstack{$\overline T_{\mathrm{2R}}$ \\ $\mathrm{(\upmu s)}$}
			&\shortstack{$\overline T_{\mathrm{2e}}$ \\ $\mathrm{(\upmu s)}$} \\
			\hline
			Transmon A & 6.01 & 0.23 & 1.2 & 95 & 45 
			& 85  \\
			Transmon B & 5.87 & 0.25 & 1.7 & 16 & 10
			& 14   \\
			Transmon C & 5.16 & 0.26 & 0.7 & 29 & 5
			& 19  \\
			Transmon D & 4.29 & 0.25 & 0.9 & 105-240 & 7-15
			& 24-26  \\
			Transmon E & 5.31 & 0.26 & 1.6 & 61 & 28
			& 29  \\
			Transmon F & 5.50 & 0.24 & 0.9 & 21 & 0.3
			& 5  \\
		\end{tabular}
	\end{ruledtabular}
\end{table}


\end{document}